\documentclass[aps,11pt]{revtex4}
\usepackage{dcolumn}
\usepackage{multirow} 
\usepackage{array}     
\usepackage{makecell}
\usepackage{graphicx}
\usepackage{amsmath}
\usepackage{amsfonts}
\usepackage{amssymb}
\usepackage{psfrag}
\usepackage{wrapfig}
\usepackage{subfigure}
\usepackage{makeidx}
\usepackage{bm}
\usepackage{epsf}
\usepackage{hyperref}
\usepackage{color}
\usepackage{cases}
\usepackage{multirow}
\usepackage{booktabs} 
\usepackage{makecell} 
\usepackage{float}

\usepackage{amsmath, amssymb, amsfonts}
\usepackage{graphicx}
\usepackage{bm}
\usepackage{xcolor}
\usepackage{ulem} 
\usepackage{hyperref}
\newcommand{\newtext}[1]{\textcolor{red}{#1}}

\makeatletter
\newcommand{\eq}[1]{Eq.(\ref{#1})}

\newcommand{\Rmnum}[1]{\uppercase\expandafter{\romannumeral #1}}
\usepackage{amsmath} 
\begin{document} 

\title{Quasinormal frequencies and greybody factors for axial perturbations of dilaton–Euler–Heisenberg de Sitter black holes}

\author{Ming Zhang$^{1}$\footnote{ zhangming@xaau.edu.cn}, Guo-Xin Chen$^{1}$\footnote{ chenguoxin@xaau.edu.cn}, Xufen Zhang$^{2}$\footnote{Corresponding author: xfzhangyzu@126.com},  Cheng-Fu Feng$^{3}$\footnote{fcfgyf@126.com}, \\  Sheng-Yuan Li$^{3}$\footnote{shengyuanli77@outlook.com},  Rui-Hong Yue$^{3}$\footnote{ rhyue@yzu.edu.cn} and De-Cheng Zou$^{4}$\footnote{dczou@jxnu.edu.cn}}

\address{
$^{1}$Faculty of Science, Xihang University, Xi'an 710077 China\\
$^{2}$Key Laboratory of Low Dimensional Quantum Structures and Quantum Control of Ministry of Education,
Synergetic Innovation Center for Quantum Effects and Applications, and Department of Physics,
Hunan Normal University, Changsha, Hunan 410081, China\\
$^{3}$Center for Gravitation and Cosmology, College of Physical Science and Technology, Yangzhou University, Yangzhou 225009, China  \\
$^{4}$School of Physics, Jiangxi Normal University, Nanchang 330022, China }

\date{\today}

\begin{abstract}
\indent

We investigate the quasinormal modes (QNMs) and greybody factors of dilaton–Euler–Heisenberg (dEH) de Sitter (dS) black holes in string-inspired Euler-Heisenberg gravity. Since the axial gravitational and electromagnetic perturbations decouple, we treat them independently.  By applying the asymptotic iteration method (AIM) alongside a sixth-order WKB approximation, we compute the quasinormal frequencies and find excellent agreement between the two approaches. 
We also find that the QNM spectra depend sensitively on the magnetic charge $Q_{\text{m}}$, cosmological constant $\Lambda$, and nonlinear coupling 
$\epsilon$, with a notable topological anomaly appearing in the electromagnetic frequency trajectories. Additionally, larger values of $Q_{\text{m}}$ or the multipole number $l$ generally suppress wave transmission, while the electromagnetic sector with $\epsilon=1$ exhibits an anomalous response.

\end{abstract}


\maketitle

\section{Introduction}
\label{intro}

Originally established in 1936\cite{Heisenberg:1936nmg}, the Euler-Heisenberg (EH) effective action offers a crucial nonlinear generalization of standard quantum electrodynamics (QED). This framework effectively captures vacuum polarization phenomena within a classical formalism, particularly in scenarios involving intense electromagnetic fields. Instead of treating the vacuum as empty space, the EH theory interprets it as a medium capable of polarization and magnetization\cite{Obukhov:2002xa}. Beyond refining classical electrodynamics in high-field scenarios, this theory has become foundational for studying nonlinear phenomena in astrophysics and cosmology.
Leveraging these distinctive properties, the first Einstein-Euler-Heisenberg (EEH) black hole solution—an anisotropic, magnetically charged configuration resembling the Reissner-Nordström metric but incorporating dyon degrees of freedom—was obtained in 1956 \cite{Yajima:2000kw}. Subsequent work has extended this to electrically charged solutions \cite{Yajima:2000kw,Ruffini:2013hia}, rotating black holes \cite{Breton:2019arv,Amaro:2022yew,Wu:2021pgf}, and frameworks within modified gravity theories \cite{Guerrero:2020uhn,Nashed:2021ctg}. Most recently, inspired by string theory and Lovelock gravity, Ref. \cite{Bakopoulos:2024hah} proposed an extension of Einstein-Maxwell-dilaton theory by coupling the dilaton field to EH electrodynamics in a novel way. Further investigations have explored particle dynamics, gravitational lensing \cite{Yasir:2025npe,Vachher:2024ezs}, and shadow properties \cite{Huang:2025jfa} around dilaton–Euler–Heisenberg (dEH)  black holes. Jiang et al. \cite{Jiang:2024njc} also analyzed the structure of geometrically thin, optically thick accretion disks in these spacetimes.

The recent detection of gravitational waves has enabled novel approaches to studying strong gravitational fields near black holes. During the ringdown phase of a binary compact star merger, the emitted gravitational waves can be modeled as the response of the remnant black hole to spacetime perturbations. At this stage, the remnant is well-described by a perturbed black hole solution. The emitted gravitational waves have characteristic decay timescales and are properly characterized by quasinormal modes (QNMs) \cite{Berti:2007,Nollert:1999}. Moreover, QNMs provide unique probes for testing the no-hair conjecture through frequency spectra \cite{Berti:2006,Berti:2007b,Isi:2019}. In the context of modified gravity theories \cite{Blazquez-Salcedo:2016enn,Aragon:2020xtm,Cano:2021myl,Zhao:2022gxl}, QNMs serve as essential tools: their behavior under parametric perturbations and background spacetime deformations strongly constrains alternative gravitational frameworks. Furthermore, investigating QNM stability helps assess whether a background spacetime is robust against small potential perturbations \cite{Jaramillo:2021,Cheung:2022,Ishibashi:2003,Yan:2020nvk,Singh:2024nvx,Media:2025xpt}.
Another crucial aspect of black hole perturbation theory is the greybody factor \cite{Konoplya:2019ppy,Konoplya:2011qq,Konoplya:2019hlu,Gogoi:2023fow,Liu:2023kxd}, which characterizes how a black hole's effective potential filters the spectrum of emitted radiation. In gravitational wave astronomy, greybody factors and QNMs form a complementary framework: QNMs determine the ringdown phase through their characteristic frequencies and damping times, while greybody factors quantify the frequency-dependent transmission probability of gravitational waves traversing the background spacetime \cite{Oshita:2023cjz,Lin:2024ubg}. 
 
These issues are particularly relevant for dEH black holes in string-inspired Euler-Heisenberg gravity. Until now, the QNMs and greybody factors of dEH black holes for test scalar and electromagnetic fields are computed in asymptotically flat \cite{Zhang:2025xqt} and de Sitter \cite{Zhang:2026nog} spacetimes. Additionally, we analyzed the polar metric perturbations \cite{Li:2026gqi} and the axial gravitational modes in the asymptotically flat case \cite{Zou:2025rbu}. 
Extending this line of research, the present paper focuses on the axial gravitational and electromagnetic perturbations in the de Sitter background.

This paper is organized as follows. In Section \ref{sec2}, we briefly review the black hole solution and derive the master equations for axial perturbations. In Section \ref{sec3}, we outline the AIM and WKB methods, and then present the numerical results for the quasinormal frequencies, analyzing their parameter dependence in details. In Section \ref{sec5}, we calculate the greybody factors using the WKB approach and discuss their physical implications. Finally, conclusions and discussions are provided in Section \ref{sec6}. Throughout this paper, we adopt geometrized units where $G = c = 1$ and a metric signature of $(-, +, +, +)$.

\section{Master equation for dEH black hole perturbations}
\label{sec2}

\subsection{Background and black hole solution}

Recently, Bakopoulos et al. \cite{Bakopoulos:2024hah} introduced a novel Einstein-Maxwell-dilaton theory that incorporates a nonlinear Euler-Heisenberg term with explicit dilaton coupling
\begin{eqnarray}
S=\frac{1}{16\pi}\int  d^4x\sqrt{-g}\Big(R-2\nabla^\mu\phi\nabla_\mu\phi-\mathfrak{V}(\phi)\Big)+\frac{1}{16\pi}\int  d^4x\sqrt{-g}{\cal L},\label{action}
\end{eqnarray}
where $R$ denotes the scalar curvature, $\phi$ is the scalar field, $ \mathfrak{V(\phi)}$ is the scalar potential and ${\cal L}$ represents the Lagrangian density describing the coupling term between scalar field and nonlinear electromagnetic field. with $ \mathfrak{V(\phi)}$ and ${\cal L}$ of the form
\begin{eqnarray}
&&\mathfrak{V(\phi)}=\frac{1}{3}\Lambda e^{-2\phi}+\frac{1}{3}\Lambda e^{2\phi}+\frac{4\Lambda}{3}=\frac{2}{3}\Lambda(\cosh(2\phi)+2),\\
&&{\cal L}=e^{-2\phi}F^2+f(\phi)\left(2\alpha F^\mu_{~\nu} F^\nu_{~\rho} F^\rho_{~\delta} F^\delta_{~\mu}-\beta F^4\right),
\end{eqnarray}
where $\Lambda$ is a positive cosmological constant, $f(\phi)$ is a coupling function, $F^2=F_{\mu\nu}F^{\mu\nu}$ and $F^4=F_{\mu\nu}F^{\mu\nu}F_{\rho\delta}F^{\rho\delta}$, where $F_{\mu\nu}$ stands for the usual field strength $F_{\mu\nu}=\partial_\mu A_{\nu}-\partial_{\nu}A_{\mu}$, where the function $f(\phi) =-\frac{1}{2}\left(3e^{-2\phi}+3e^{2\phi}+4\right)$. If one sets $\alpha=\beta=0$, the model will reduce to the standard Einstein-Maxwell-dilaton theory. 
  Varying the Einstein equation from the action \eqref{action}
  \begin{eqnarray}
G_{\mu\nu}=2{\partial _\mu}\phi{\partial_ \nu}\phi-g_{\mu\nu}{\partial }^\mu\phi {\partial }_\mu \phi+2T_{\mu\nu}^{\phi}+T_{\mu\nu}^{A}-\frac{1}{2}g_{\mu\nu}\mathfrak{V(\phi)},\label{eqG}
 \end{eqnarray}
 where $G_{\mu\nu}$ is the Einstein tensor and the energy-momentum tensor takes the form
\begin{eqnarray}
&&G_{\mu\nu}=R_{\mu\nu}-\frac{1}{2}g_{\mu\nu}R, \\
&&T_{\mu\nu}^{\phi}=e^{-2\phi}(F^{\alpha}_\mu F_{\nu \alpha}-\frac{1}{4}g_{\mu\nu }F^2),\nonumber\\
&&T_{\mu\nu}^{A}=f(\phi)( 8\alpha F^\alpha _\mu F^\beta_\nu F^\eta_\alpha F_{\beta\eta} -\alpha g_{\mu\nu} F^\alpha_\beta F^\beta _\gamma F^\gamma _\delta F^\delta_\alpha -4\beta F^\xi_\mu F_{\nu\xi}F^2+\frac{1}{2}g_{\mu\nu}\beta F^4),
\end{eqnarray}
the scalar equation is given by
\begin{eqnarray}
\Box  \phi +\frac{1}{2}e^{-2\phi}F^2-\frac{df(\phi)}{d\phi}\left(\frac{\alpha}{2} F^\mu_{~\nu} F^\nu_{~\gamma} F^\gamma_{~\delta} F^\delta_{~\mu}-\frac{\beta}{4} F^4\right)-\frac{1}{4}\frac{d\mathfrak{V(\phi)}}{d \phi}=0,\label{eqKG}
\end{eqnarray}
the Maxwell equation leads to
\begin{eqnarray}
&&{\partial_ \mu}\Big[\sqrt{-g}\left(e^{-2\phi}F^{\mu\nu}+f(\phi)(4\alpha F^\mu_{~\kappa} F^\kappa_{~\lambda} F^{\nu\lambda}-2\beta F^2 F^{\mu\nu})\right)\Big]=0\label{eqMaxwell}
\end{eqnarray}

For convenience, we define $\epsilon \equiv \alpha-\beta$,
The magnetically charged black hole solution in de sitter spacetime can be written as~\cite{Bakopoulos:2024hah, Zhang:2026nog}
\begin{eqnarray}
ds^2 & =& -A(r)dt^2 + \frac{1}{B(r)} dr^2 + r^2 (d\theta^2 + \sin^2 \theta d\varphi^2),\label{metric}\\
A(r)&=&1-\frac{4 M^2}{Q_{\text{m}}^2+\sqrt{Q_{\text{m}}^4+4 M^2 r^2}}-\frac{2\epsilon Q_{\text{m}}^4}{r^6}-\frac{1}{3}\Lambda r^2,\nonumber\\
B(r)&=&1
- \frac{Q_{m}^{4} + 4M^{2}r^{2}}{r^2(Q_{m}^{2} + \sqrt{Q_{m}^{4} + 4M^{2}r^{2}})} +\frac{Q_{\text{m}}^4}{4M^2r^2}
- \frac{\epsilon Q_{m}^{4}(Q_{m}^{4} + 4M^{2}r^{2})}{2M^2r^{8}}-\Lambda\left(\frac{r^2}{3}+\frac{Q_{\text{m}}^4}{12M^2}\right),\nonumber\\
\phi (r)&=&-\frac{1}{2}\ln \left(\frac{\sqrt{Q_{\text{m}}^4+4 M^2 r^2}-Q_{\text{m}}^2}{\sqrt{Q_{\text{m}}^4+4 M^2 r^2}+Q_{\text{m}}^2}\right)\label{solution}.
\end{eqnarray}
Notice that the magnetic gauge potential is
$\boldsymbol{A}_\mu=(0,0,0,Q_{\rm m}\cos\theta)$,
$M$ and $Q_{\rm m}$ denote the black hole mass and magnetic charge, respectively.

\begin{figure}[H]
\centering
\subfigure[$\epsilon=-1$ and $\Lambda=0.05$]
{\label{fig1-1} 
\includegraphics[width=2.8in]{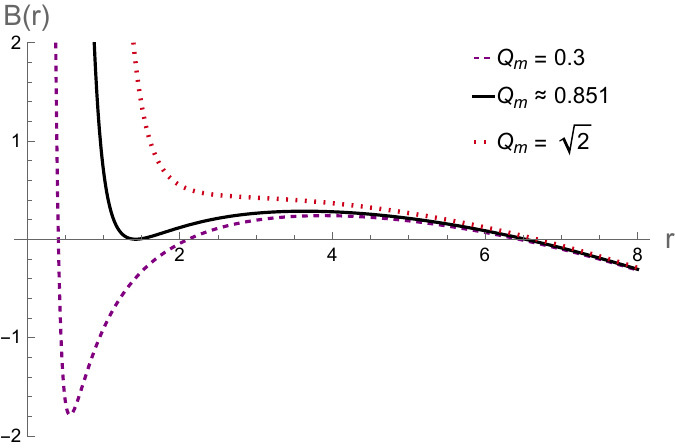}}
\hspace{1em}
\subfigure[$\epsilon=-1$ and $Q_{\text{m}}=0.3$]
{\label{fig1-2} 
\includegraphics[width=2.8in]{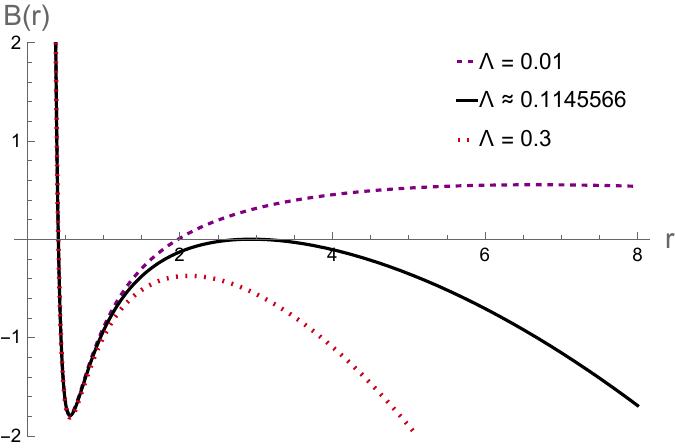}}
\caption{The metric function $B(r)$ as a function of the radial coordinate $r$ with $M=1$}\label{fig1}
\end{figure}
\begin{figure}[H]
\centering
\subfigure[$\epsilon=0$ and $\Lambda=0.05$]
{\label{fig2-1} 
\includegraphics[width=2.8in]{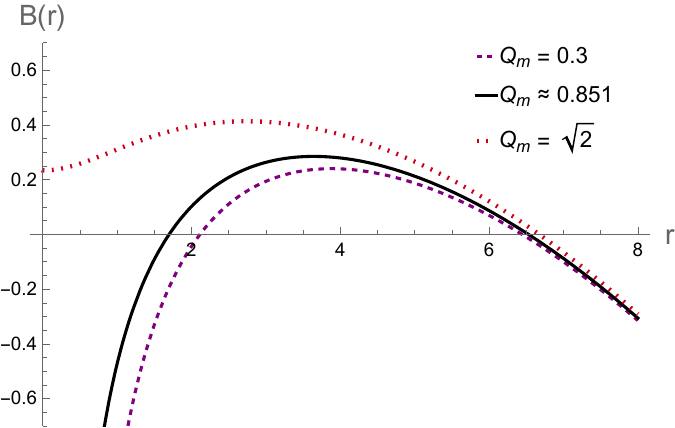}}
\hspace{1em}
\subfigure[$\epsilon=1$ and $\Lambda=0.05$]
{\label{fig2-2} 
\includegraphics[width=2.8in]{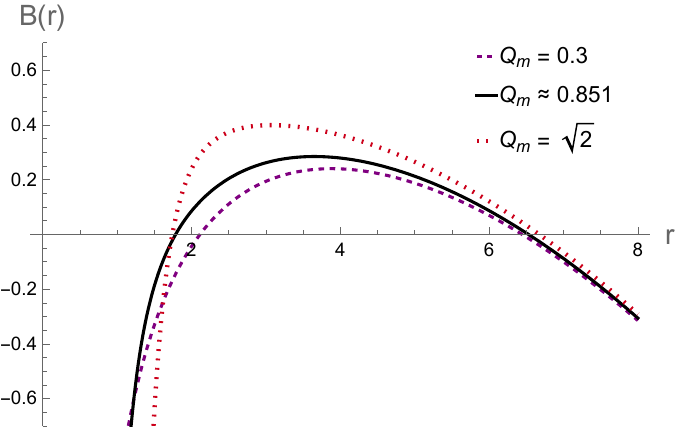}}
\caption{The metric function $B(r)$ as a function of the radial coordinate $r$  with $M=1$}\label{fig2}
\end{figure}

The parameter $\epsilon$ significantly influences the horizon structure of the spacetime, yielding qualitatively distinct geometrical configurations across different parameter spaces, as illustrated in Figs. \ref{fig1} and \ref{fig2}.

For $\epsilon < 0$, the spacetime exhibits a rich horizon structure that can accommodate up to three distinct horizons: a Cauchy (inner) horizon, a black hole event horizon, and a cosmological horizon. The left panel of Fig. \ref{fig1} demonstrates the effect of the magnetic charge $Q_{\text{m}}$ for a fixed cosmological constant $\Lambda$. As $Q_{\text{m}}$ increases, the Cauchy horizon and the event horizon gradually approach each other, eventually merging at a critical value ($Q_{\text{m}} \approx 0.851$) to form an extremal black hole. When the charge is increased further (e.g., $Q_{\text{m}} = \sqrt{2}$), the black hole horizons vanish entirely. In this regime, the central singularity is no longer hidden by an event horizon, resulting in a naked singularity encompassed only by the cosmological horizon. The right panel of Fig. \ref{fig1} illustrates the influence of the cosmological constant $\Lambda$ for a fixed magnetic charge. As $\Lambda$ increases, the black hole event horizon expands outward while the cosmological horizon shrinks inward until the two coincide ($\Lambda \approx 0.1145$), corresponding to the Nariai limit. For even larger values of $\Lambda$, the black hole and cosmological horizons disappear, plunging the spacetime into another naked singularity phase.

Conversely, the horizon structures for $\epsilon = 0$ and $\epsilon > 0$ are considerably less complex, as displayed in Fig. \ref{fig2}. In these regimes, the spacetime typically possesses a standard two-horizon structure (an event horizon and a cosmological horizon) within the considered parameter ranges. For $\epsilon = 0$ (the left panel of Fig. \ref{fig2}), the peak of the metric function $B(r)$ decreases with an increasing $Q_{\text{m}}$, showing a trend toward an extremal configuration near $Q_{\text{m}} = \sqrt{2}$. However, for $\epsilon = 1$ (the right panel of Fig. \ref{fig2}), the nonlinear electrodynamic effects significantly suppress the influence of the magnetic charge. Consequently, the qualitative horizon structure remains remarkably stable and essentially unchanged throughout the examined parameter space.

\subsection{Axial perturbations}

The study of black hole perturbations was initially conducted by Regge and Wheeler \cite{Bonanno:2025ffy} who focused on odd-parity spherical harmonic modes, with subsequent extension to even-parity modes by Zerilli \cite{Zerilli:1970se}.
This work focuses specifically on odd-parity perturbations to simplify the theoretical framework. We regard the perturbed metric as a sum of the unperturbed background metric $\bar{g}_{\mu\nu}$ and the actual perturbation $\delta{g}_{\mu\nu}$
\begin{eqnarray}
  g_{\mu\nu}=\bar{g}_{\mu\nu}+\delta{g}_{\mu\nu}, \label{pertmetric}
\end{eqnarray}
Working within the Regge-Wheeler gauge formalism\cite{Regge:1957td}, we decompose the metric fluctuations into tensor spherical harmonics. For the odd-parity (axial) sector, the gravitational perturbation is characterized by two radial functions, $h_0(r)$ and $h_1(r)$, leading to the following metric ansatz:
\begin{eqnarray}
\delta{g}_{\mu\nu}=
\sum_{l,m} e^{-i\omega t}\begin{bmatrix}
0 & 0 &-\frac{h_0(r){\partial _\varphi}}{\sin\theta} & h_0(r)\sin\theta\partial_{\theta} \\
0 & 0 &-\frac{h_1(r){\partial _\varphi}}{\sin\theta} & h_1(r)\sin\theta\partial_{\theta} \\
-\frac{h_0(r){\partial _\varphi}}{\sin\theta} & -\frac{h_1(r){\partial _\varphi}}{\sin\theta} & 0 & 0 \\
h_0(r)\sin\theta\partial_{\theta} & h_1(r)\sin\theta\partial_{\theta} & 0 & 0
\end{bmatrix}Y_{lm}
, \label{eqodd}
\end{eqnarray}
where the spherical harmonics can be replaced by Legendre polynomials by setting the azimuthal number $m=0$ without loss of generality, i.e., $Y_{lm}|_{m=0}=\sqrt{\frac{2l+1}{4\pi}}P_{l}(\cos\theta)$, because the background metric is spherically symmetric. Similarly, we write the vector perturbations as
\begin{eqnarray}
\quad  A_{\mu}=\bar{A}_{\mu}+\delta{A}_{\mu},\label{pertmetric1}
\end{eqnarray}
where $\bar{A}_{\mu}$ represents the background electromagnetic field, and $\delta{A}_{\mu}$ denotes the corresponding perturbations. The axial gauge transformation will be analyzed, with the gauge vector characterized by the specific form: 
\begin{eqnarray}\label{vectorpert}
\delta{A}_{\mu}=\sum_{l,m} e^{-i\omega t}\Big[0,0,-u_3(r)\frac{\partial_{\varphi}}{\sin\theta},u_3(r)\sin\theta\partial_{\theta}\Big]Y_{lm}.
\end{eqnarray}
Substituting the perturbed metric and vector potential (Eqs.~\eqref{pertmetric}, \eqref{eqodd}, \eqref{pertmetric1}, and \eqref{vectorpert}) into the field equation \eqref{eqG}, the non-zero components of the first-order perturbed gravitational field equation can be written as:
\begin{eqnarray}
E^{(g)}_{tt}&=&E_{rr}^{(g)}=E_{t\theta}^{(g)}= \Big[r^4-2\epsilon Q_{\text{m}}^2\left(4
   e^{2 \phi}+3 e^{4 \phi}+3\right)\Big]u_3,\label{maxpert1}\\
E^{(g)}_{r\theta}&=&\Big[r^4-2\epsilon Q_{\text{m}}^2\left(4
   e^{2 \phi}+3 e^{4 \phi}+3\right)\Big]u_3',\label{maxpert2}\\
E^{(g)}_{\theta \theta}&=&E_{\varphi\varphi}^{(g)}=\Big[r^4-6\epsilon Q_{\text{m}}^2\left(4
   e^{2 \phi}+3 e^{4 \phi}+3\right)\Big]u_3,\label{maxpert3}\\
E_{t\varphi}^{(g)}&=&-3r^2B h_0A'^2+3r^2A \Big[h_0A' B'+B\left( i\omega h_1A'+A' h_0'+2h_0 A'' \right)\Big]\nonumber\\
&&+A^2\Big[2h_0\left(3l^2+3l-6+4r^2 \Lambda+2r^2  \Lambda \cosh(\phi)+3r^2 {\cal \bar{L}}+6r B' +6B(1+r^2 \phi '^2)\right)\nonumber\\
&&-3i r \left(r B'(\omega h_1-i h_0')+2B(2\omega h_1+r \omega h_1'-i r h_0'')\right)\Big],\label{E14}\\
E_{r\varphi}^{(g)}&=&2 A^2 h_1\left(3l^2+3l-6+4r^2 \Lambda+2r^2  \Lambda \cosh(\phi)+3r^2 {\cal \bar{L}}+3r( B' +2Br^2 \phi '^2)\right)\nonumber\\
&&-3r^2 B h_1 A'^2-3r A \Big[4i\omega h_0-2i r\omega h_0'+h_1\left(r(2\omega^2-A' B')-2B(A'+rA'')\right)\Big],\label{E24}\\
E_{\theta\varphi}^{(g)}&=&2i\omega h_0+ A h_1 B'+B(h_1 A' + 2 A h_1').\label{E34}
\end{eqnarray}
From Eqs.~\eqref{maxpert1}, \eqref{maxpert2}, and \eqref{maxpert3}, we obtain the perturbation function $u_3(r)=0$. Moreover, this function does not appear in the gravitational perturbation equations (\eqref{E14}, \eqref{E24}, and \eqref{E34}), which clearly demonstrates that the axial gravitational perturbations are decoupled from the electromagnetic ones. This decoupling between the metric and gauge field sectors is consistent with previous studies on magnetic black holes~\cite{Nomura:2020tpc,Meng:2022oxg,Daghigh:2021psm}, allowing us to treat the gravitational sector independently.

We now focus on the axial electromagnetic perturbations.
We restrict our analysis to the axial sector, as polar perturbations generally involve a more complicated coupling among the metric, electromagnetic, and dilaton fields; see Ref.~\cite{Li:2026gqi} for a recent study of the coupled metric--dilaton sector.
Substituting the perturbed metric and vector potential (Eqs.~\eqref{pertmetric}, \eqref{eqodd}, \eqref{pertmetric1}, and \eqref{vectorpert}) into the Maxwell equation~\eqref{eqMaxwell}, we obtain the corresponding perturbed equations.
\begin{eqnarray}
E^{(e)}_{t}&=&h_0\Big[2\epsilon Q_{\text{m}}^2\left(4
   e^{2 \phi}+3 e^{4 \phi}+3\right)-r^4\Big],\label{app1}\\
E^{(e)}_{r}&=&h_1\Big[2\epsilon Q_{\text{m}}^2\left(4
   e^{2 \phi}+3 e^{4 \phi}+3\right)-r^4\Big],\label{app2}\\
E^{(e)}_{\theta}&=&h'_1\Big[r^4-2\epsilon Q_{\text{m}}^2\left(4
   e^{2 \phi}+3 e^{4 \phi}+3\right)\Big],\label{app3}\\
E^{(e)}_{\varphi}&=&2\Bigg[\Big(r^2\omega^2\Big[r^4-2\epsilon Q_{\text{m}}^2\left(4
   e^{2 \phi}+3 e^{4 \phi}+3\right)\Big]+l(1+l)A\Big[6\epsilon Q_{\text{m}}^2\left(4
   e^{2 \phi}+3 e^{4 \phi}+3\right)\nonumber\\
   &&-r^4\Big]\Bigg]u_3+r^6\Bigg[u'_3\Big(AB'+B(A'-4A\phi')\Big)+2ABu''_3\Bigg]-4e^{2\phi}r\epsilon Q^2_m\Bigg[u'_3\Big((2+3\cosh(2\phi))\nonumber\\
   &&\times \Big(rBA'+A(rB'-8B)\Big)+12rAB\sinh(2\phi)\phi'+2rAB(2+3\cosh(2\phi))u''_3\Bigg].
\label{app4}
\end{eqnarray}
From these equations, it can be seen that the gravitational perturbation functions $h_0(r)$ and $h_1(r)$ must vanish and there only exist single perturbation equation for perturbed electromagnetic field. 

\subsubsection{Axial gravitational perturbation}

It is well known that gravitational perturbations are central to black hole physics because they directly probe spacetime geometry, align with observational priorities in gravitational wave astronomy, and reflect the black hole’s intrinsic properties. In the subsequent section, we only focus on the axial gravitational perturbation.
From Eqs.\eqref{E14}, \eqref{E24} and \eqref{E34}, it's easily verified that only two of the above three
equations are independent.  Considering the component $E_{\theta\varphi}^{(g)}$ \eqref{E34}, we have 
\begin{eqnarray}
h_0=\frac{i}{2\omega}\Big[B(h_1 A'+2A h_1') +A h_1 B'\Big]. \label{h0}
\end{eqnarray}
Substituting \eq{h0} into the component $E_{r\varphi}^{(g)}$ \eqref{E24}, we can eliminate $h_0(r)$ and $h_0'(r)$, and then obtain a single second order differential equation for $h_1(r)$.
In order to cast this master equation into the standard Schr\"odinger form, we further define the function $\Psi_G(r)$ with
\begin{eqnarray}
h_1(r)=C_0(r)*\Psi_G(r).
\end{eqnarray}

We assume the function $C_0(r)$ taking the following form
\begin{eqnarray}
C_0(r)=\frac{r}{\sqrt{A(r)B(r)}},
\end{eqnarray}
and then obtain the final perturbed equation
\begin{eqnarray}
 \frac{d^2\Psi_G(r_*)}{d r^2_{*}}+\Big[\omega^2-V_G(r)\Big]\Psi_G(r_*)=0,\label{perteq}
\end{eqnarray}
where $r_*$ represents the tortoise coordinate with
\begin{eqnarray}
d{r_*} = \frac{1}{\sqrt{AB}}dr,
\end{eqnarray}
and the effective potential $V(r)$ is
\begin{eqnarray}
V_G(r)&=&\frac{1}{6} \Big[ \frac{1}{r} \Big(3 A'(r) \left( B(r) + r B'(r) \right) \Big)-\frac{3 B(r) A'(r)^2}{A(r)} +  \nonumber\\
&&+\frac{A(r)}{r^8} \Big( 12 e^{-2\phi(r)} r^4 Q_{\text{m}}^2 - 24 \epsilon (2 + 3 \cosh(2\phi(r))) Q_{\text{m}}^4  \nonumber\\
&&+ r^6 \left( 6(l^2+l-2) + 8 r^2 \Lambda + 4 r^2 \Lambda \cosh(2\phi(r)) + 3 r B'(r) \right)\nonumber\\
&&+ 12 B(r) \left( r^6 + r^8 \phi'(r)^2 \right) \Big) 
+ 6 B(r) A''(r) \Big] .\label{V1}
\end{eqnarray}
with the Lagrangian density ${\cal \bar{L}}$. Its detailed form with black hole solutions \eqref{solution} can be written as 
\begin{eqnarray}
{\cal \bar{L}}(\phi,F)=\frac{2 Q_\mathrm{m}^2 e^{-2\phi (r)}}{r^4}-\frac{4 \epsilon Q_{\text{m}}^4 (3 \cosh (2 \phi (r))+2)}{r^8}\label{LFF}
\end{eqnarray}
We should mention that perturbation of the energy-momentum tensor of electromagnetic part can not be ignored. The similar phenomenon also appears in Ref.\cite{Wu:2018xza}.

\begin{figure}[H]
\centering
\subfigure[$\epsilon=-1$]{\label{fig3-1} 
\includegraphics[width=2in]{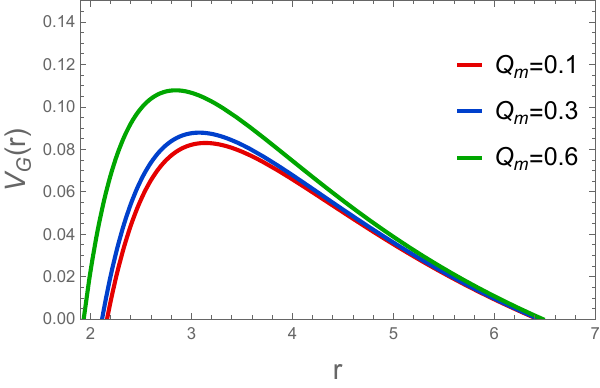}}
\hfill
\subfigure[$\epsilon=0$]{\label{fig3-2} 
\includegraphics[width=2in]{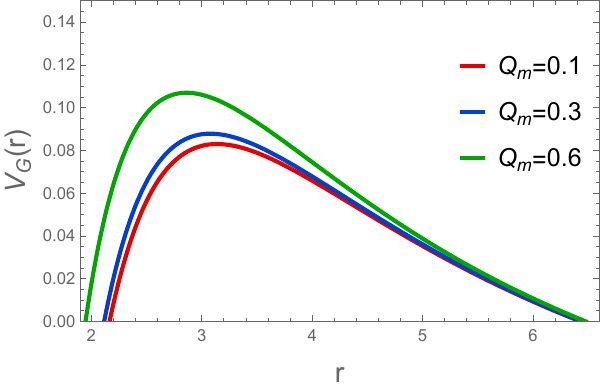}}
\hfill
\subfigure[$\epsilon=1$]{\label{fig3-3} 
\includegraphics[width=2in]{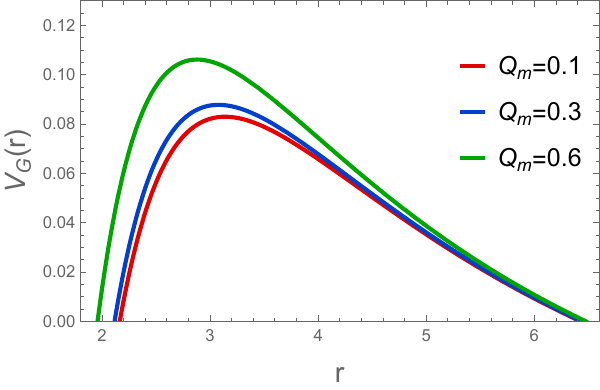}}
\caption{The effective potential $V_{\mathrm{G}}(r)$ for gravitational field  perturbation ($l=2$) as a function of the radial coordinate $r$ for different values of magnetic charge $Q_{\mathrm{m}}$. In all cases, we set $M=1$ and $\Lambda=0.05$. Panels (a), (b), and (c) correspond to $\epsilon=-1$, $\epsilon=0$, and $\epsilon=1$, respectively. As $Q_{\mathrm{m}}$ decreases, the peak of the potential barrier shifts outward and its height is significantly suppressed.}\label{fig3}
\end{figure}
\begin{figure}[H]
\centering
\subfigure[$\epsilon=-1$]{\label{fig4-1} 
\includegraphics[width=2in]{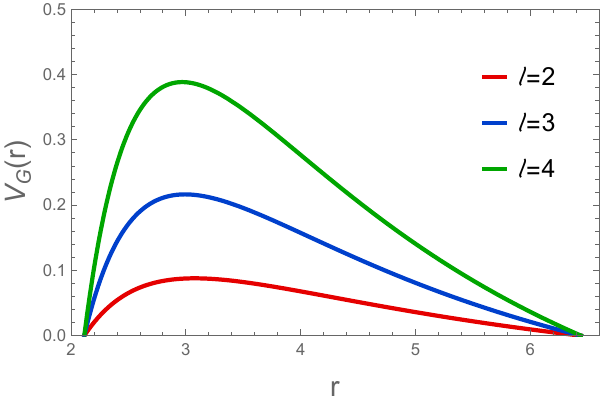}}
\hfill
\subfigure[$\epsilon=0$]{\label{fig4-2} 
\includegraphics[width=2in]{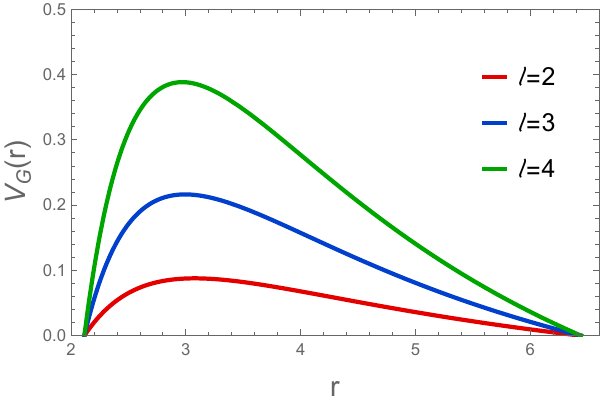}}
\hfill
\subfigure[$\epsilon=1$]{\label{fig4-3} 
\includegraphics[width=2in]{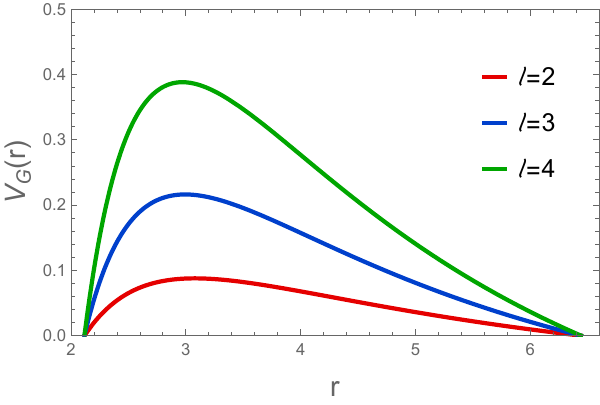}}
\caption{The effective potential $V_{\mathrm{G}}(r)$ for gravitational field perturbations (with $Q_{\mathrm{m}}=0.3$) as a function of the radial coordinate $r$, for different values of the angular quantum number $l$. In all cases, we set $M=1$ and $\Lambda=0.05$. Panels (a), (b), and (c) correspond to $\epsilon=-1$, $\epsilon=0$, and $\epsilon=1$, respectively. As $l$ increases, the peak of the potential barrier shifts to the right and its height increases significantly.}\label{fig4}
\end{figure}

Note that the solutions \eqref{solution} and potential \eqref{V1} are invariant under the following rescaling: $r/M\rightarrow r$, $Q_{\text{m}} /M\rightarrow Q_{\text{m}}$, and  $\epsilon/ M^2\rightarrow \epsilon$. 
For a better analysis of the potential function's behavior, 
we will set $M=1$ throughout the paper and leave $\epsilon$ and $Q_{\text{m}}$ free without loss of generality. The effective potentials $V_G(r)$ are plotted in Figs.\ref{fig3} and \ref{fig4}, respectively. It is seen that the height of the effective potential increases as $Q_{\text{m}}$ increases. Moreover, the potential also increases for an increase the multipole moment $l$ in both cases.It's worthy to point out that the effective potentials are always positive, indicating that the system is stable under the axial gravitational field perturbation.

\subsubsection{Axial electromagnetic perturbation}

\begin{figure}[H]
\centering
\subfigure[$\epsilon=-1$]{\label{fig5-1} 
\includegraphics[width=2in]{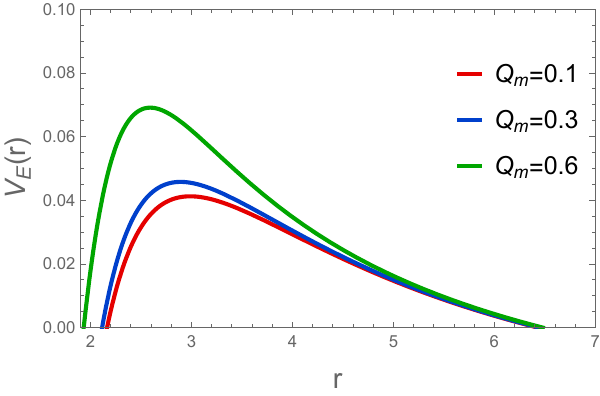}}
\hfill
\subfigure[$\epsilon=0$]{\label{fig5-2} 
\includegraphics[width=2in]{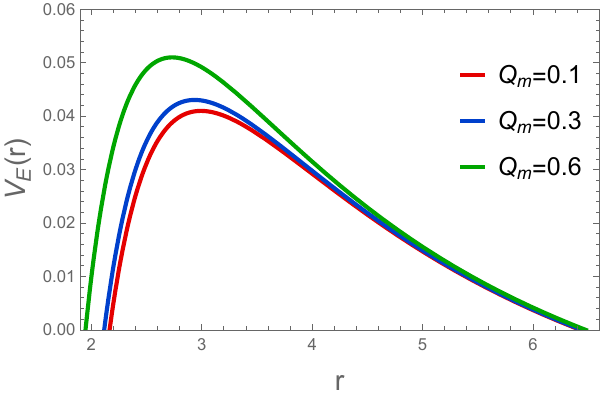}}
\hfill
\subfigure[$\epsilon=1$]{\label{fig5-3} 
\includegraphics[width=2in]{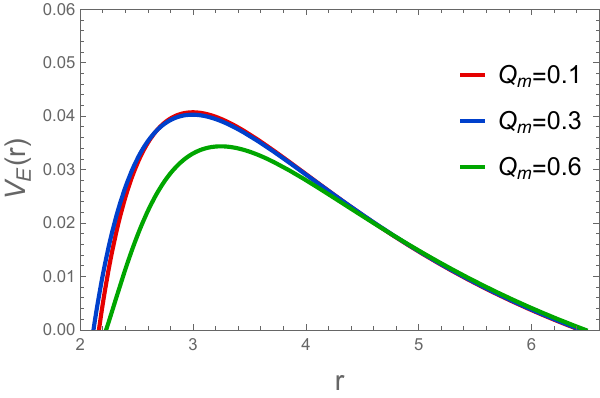}}
\caption{The effective potential $V_{\mathrm{E}}(r)$ for electromagnetic field  perturbation ($l=1$) as a function of the radial coordinate $r$ for different values of magnetic charge $Q_{\mathrm{m}}$. In all cases, we set $M=1$ and $\Lambda=0.05$. Panels (a), (b), and (c) correspond to $\epsilon=-1$, $\epsilon=0$, and $\epsilon=1$, respectively.}\label{fig5}
\end{figure}

\begin{figure}[H]
\centering
\subfigure[$\epsilon=-1$]{\label{fig6-1} 
\includegraphics[width=2in]{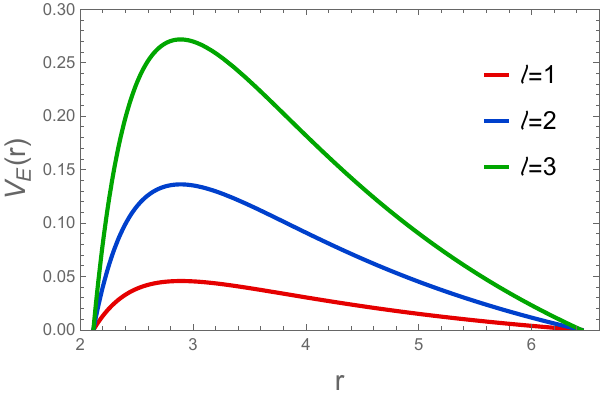}}
\hfill
\subfigure[$\epsilon=0$]{\label{fig6-2} 
\includegraphics[width=2in]{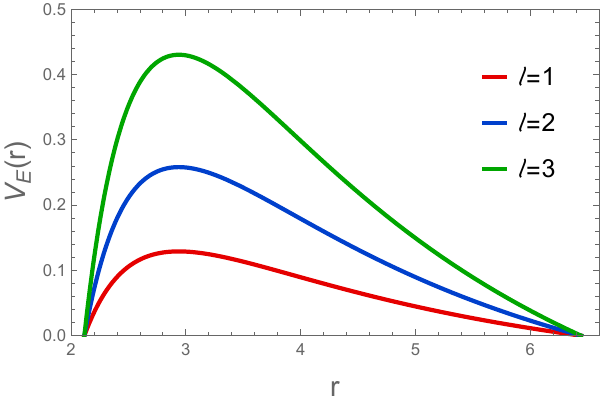}}
\hfill
\subfigure[$\epsilon=1$]{\label{fig6-3} 
\includegraphics[width=2in]{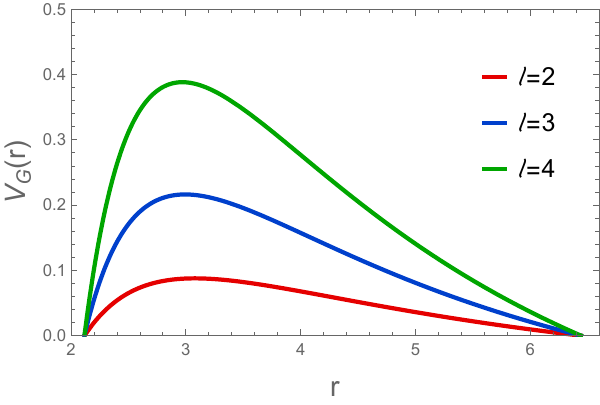}}
\caption{The effective potential $V_{\mathrm{E}}(r)$ for electromagnetic field perturbations (with $Q_{\mathrm{m}}=0.3$) as a function of the radial coordinate $r$, for different values of the angular quantum number $l$. In all cases, we set $M=1$ and $\Lambda=0.05$. Panels (a), (b), and (c) correspond to $\epsilon=-1$, $\epsilon=0$, and $\epsilon=1$, respectively. As $l$ increases, the peak of the potential barrier shifts to the right and its height increases significantly.}\label{fig6}
\end{figure}

In order to cast this master equation \eqref{app4} into the standard Schr\"odinger form, we further define the function $u_3(r)$ with
\begin{eqnarray}
u_3(r)=C_1(r)*\Psi_E(r).
\end{eqnarray}
We assume the function $C_1(r)$ taking the following form
\begin{eqnarray}
C_1(r)=\frac{r^2 e^{\phi(r)}  }{\sqrt{ r^4 - 2 \left(3 + 4 e^{2\phi(r)} + 3 e^{4\phi(r)}\right) \epsilon Q_{\text{m}}^2 }},
\end{eqnarray}
and then obtain the final perturbed equation
\begin{eqnarray}
 \frac{d^2\Psi_E(r_*)}{d r^2_{*}}+\Big[\omega^2-V_E(r)\Big]\Psi_E(r_*)=0,
\end{eqnarray}
where 
 \begin{eqnarray}
V_E(r)=&&-\frac{1}{2 \left( r^5 - 2 r \epsilon Q_{\text{m}}^2 \left(4 e^{2 \phi(r)} + 3 e^{4 \phi(r)} + 3\right) \right)^2} \nonumber\\
&& \times \Bigg[ -2 r \epsilon B(r) Q_{\text{m}}^2 A'(r) \left(-3 r \left(e^{4 \phi(r)}-1\right) \phi'(r) + 8 e^{2 \phi(r)} + 6 e^{4 \phi(r)} + 6\right) \nonumber\\
&& \times \left(r^4 - 2 \epsilon Q_{\text{m}}^2 \left(4 e^{2 \phi(r)} + 3 e^{4 \phi(r)} + 3\right)\right) + r^6 B(r) A'(r) \phi'(r) \left(r^4 - 2 \epsilon Q_{\text{m}}^2 \left(4 e^{2 \phi(r)} + 3 e^{4 \phi(r)} + 3\right)\right) \nonumber\\
&& - A(r) \Bigg( 4 \epsilon^2 Q_{\text{m}}^4 \bigg( \left(4 e^{2 \phi(r)} + 3 e^{4 \phi(r)} + 3\right) \Big( r B'(r) \Big(3 r \left(e^{4 \phi(r)}-1\right) \phi'(r) \nonumber\\
&& - 2 \left(4 e^{2 \phi(r)} + 3 e^{4 \phi(r)} + 3\right)\Big) + 6 l (l+1) \left(4 e^{2 \phi(r)} + 3 e^{4 \phi(r)} + 3\right) \Big) \nonumber\\
&& + 6 B(r) \Big( \left(4 e^{2 \phi(r)} + 3 e^{4 \phi(r)} + 3\right) \left(r^2 \left(e^{4 \phi(r)}-1\right) \phi''(r) + 8 e^{2 \phi(r)} + 6 e^{4 \phi(r)} + 6\right) \nonumber\\
&& + r^2 \left(8 e^{2 \phi(r)} + 18 e^{4 \phi(r)} + 8 e^{6 \phi(r)} + 3 e^{8 \phi(r)} + 3\right) \phi'(r)^2 \nonumber\\
&&- 4 r \left(-4 e^{2 \phi(r)} + 4 e^{6 \phi(r)} + 3 e^{8 \phi(r)} - 3\right) \phi'(r) \Big) \bigg) \nonumber\\
&&- 4 r^4 \epsilon Q_{\text{m}}^2 \Bigg( -r B'(r) \left(r \left(2 e^{2 \phi(r)} + 3\right) \phi'(r) + 4 e^{2 \phi(r)} + 3 e^{4 \phi(r)} + 3\right) \nonumber\\
&& + 2 B(r) \Big( -r^2 \left(2 e^{2 \phi(r)} + 3\right) \phi''(r) + r^2 \left(4 e^{2 \phi(r)} + 9 e^{4 \phi(r)} + 3\right) \phi'(r)^2 \nonumber\\
&& - 2 r \left(4 e^{2 \phi(r)} + 9 e^{4 \phi(r)} - 3\right) \phi'(r) + 5 \left(4 e^{2 \phi(r)} + 3 e^{4 \phi(r)} + 3\right) \Big) \nonumber\\
&& + 4 l (l+1) \left(4 e^{2 \phi(r)} + 3 e^{4 \phi(r)} + 3\right) \Bigg) \nonumber\\
&&\ + r^8 \left(r^2 (-B'(r)) \phi'(r) + 2 r^2 B(r) \left(\phi'(r)^2 - \phi''(r)\right) + 2 l (l+1)\right) \Bigg] .
\end{eqnarray}
The radial profile of the effective potential $V_E(r)$ exhibits distinct characteristics for different combinations of the magnetic charge $Q_{\mathrm{m}}$ and angular quantum number $l$, as depicted in Figs.~\ref{fig5} and~\ref{fig6}. Crucially, for all parameter sets considered, the potential maintains positive definiteness throughout the exterior spacetime region ($r_{\mathrm{h}}<r<r_{\mathrm{c}}$). This persistent positivity, particularly the absence of any negative potential well, provides strong evidence for the black hole's structural stability against electromagnetic field perturbations. Furthermore, quantitative analysis reveals that, for $\epsilon=1$, the height of the potential barrier is anomalously suppressed with increasing $Q_{\mathrm{m}}$, indicating a distinct geometric dependence induced by the nonlinear electrodynamics, in contrast to the monotonically enhancing behavior observed for $\epsilon\le 0$.

\section{Quasinormal Mode frequencies}
\label{sec3}

In order to rigorously compute the quasinormal spectra, we utilize a combination of the Asymptotic Iteration Method (AIM)~\cite{Cho:2009cj,Cho:2011sf,Devi:2026tsr,Singh:2025tvk} and the WKB approximation~\cite{Iyer:1986np,Iyer:1986nq, Konoplya:2003ii,Singh:2025fuv}. By applying these two independent numerical techniques, we can ensure the reliability of our frequency data through mutual verification.

\subsection{Asymptotic iteration method}

In this section, we will show the main steps of the AIM method. Firstly, we rewrite axial metric perturbed equation \eqref{perteq} in terms of $u=1/r$
\begin{eqnarray}
&&\left( 2 u^3 A(u) B(u) - \frac{1}{2} u^2 \left( -u^2 B(u) A'(u) - u^2 A(u) B'(u) \right) \right) \Psi'(u)
\quad + \frac{1}{6} \Psi(u) \Bigg( 6 \omega^2 + \frac{3 u^4 B(u) A'(u)^2}{A(u)}  \nonumber\\
&&+ 3 u^3 A'(u) \left( B(u) - u B'(u) \right) 
 + u^8 A(u) \Bigg( -\frac{12 e^{-2\phi(u)} Q_{\text{m}}^2}{u^4} + 24 \epsilon (2 + 3 \cosh(2\phi(u))) Q_{\text{m}}^4 \nonumber\\
&& - \frac{ 6}{u^6}\Big((l^2+l-2) + \frac{8\Lambda}{u^2} + \frac{4\Lambda \cosh(2\phi(u))}{u^2} - 3 u B'(u) \Big) 
- 12 B(u) \left( \frac{1}{u^6} + \frac{r_h^2 \phi'(u)^2}{u^4} \right) \Bigg) \nonumber\\
&& - 6 B(u) \left( 2 u^3 A'(u) + u^4 A''(u) \right) \Bigg) 
 + u^4 A(u) B(u) \Psi''(u)=0
\end{eqnarray}
To scale out the divergent behavior at the cosmological and event horizons, we define
\begin{eqnarray}
  \psi(u)=e^{i\omega r_*}(u-u_1)^{-\frac{i\omega}{\kappa_1}} \chi(u)\label{eqschi}.
\end{eqnarray}
where $\chi(u)$ is a finite and convergent function. To establish the proper scaling behavior satisfying the quasinormal mode boundary conditions, we define
\begin{eqnarray}
e^{i\omega r_*}=\prod_{j}(u-u_j)^{i\omega/\kappa_j},
\end{eqnarray}
where $\kappa_j$ denotes the surface gravity evaluated at the coordinate $u_j$, defined by the condition $B(u=u_j)=0$. This scaling transformation regularizes the solution by removing the divergent behavior near the $u_j$ boundary and simultaneously enforces the boundary conditions. We introduce the surface gravity at the event horizon:
\begin{eqnarray}
\kappa_h=\frac{1}{2}\frac{ d\sqrt{A(r)B(r)}}{d r}\mid_{r=r_h}
\end{eqnarray}
Here, $r_h$ and $r_c$ denote the radii of the black-hole event horizon and the cosmological horizon, respectively. Both are positive roots of the metric function $B(r)=0$ and satisfy $r_h<r_c$. If a Cauchy horizon is present, its radius $r_-$ satisfies $r_-<r_h<r_c$. A similar evaluation of the surface gravity and horizon boundary conditions in the context of quasinormal modes for black holes with a cosmological constant can be found in Ref.~\cite{Singh:2025fuv}.
Substituting Eq.~\eqref{eqschi} into the $u$-domain perturbation equation, we obtain
\begin{eqnarray}
\chi''=\lambda_0(u)\chi'+s_0(u)\chi \label{eqschieq},
\end{eqnarray} 
where 
\begin{eqnarray}
  \lambda_0(u)=\frac{1}{2}\Big[-\frac{4}{u}-\frac{4i r_c\omega}{\kappa_c-\kappa_c r_c u}-\frac{4i r_h \omega}{\kappa_h-\kappa_h r_h u}-\frac{A'(u)}{A(u)}-\frac{B'(u)}{B(u)}\Big],
\end{eqnarray}
and
\begin{eqnarray}
s_0(u)&=&\frac{1}{6B(u)}
\Bigg(
\frac{2\Lambda}{u^4}
\left(4+e^{-2\phi(u)}+e^{2\phi(u)}\right)
+\frac{6\left[l(l+1)-2\right]}{u^2}
+12e^{-2\phi(u)}Q_{\mathrm{m}}^2
\nonumber\\
&&
-12e^{-2\phi(u)}
\left(3+4e^{2\phi(u)}+3e^{4\phi(u)}\right)
\epsilon u^4Q_{\mathrm{m}}^4
-\frac{3B'(u)}{u}
+\frac{1}{A(u)}
\left(
3A'(u)B'(u)-\frac{6\omega^2}{u^4}
\right)
\Bigg)
\nonumber\\
&&
+\frac{1}{2}
\Bigg(
\frac{4}{u^2}
-\frac{2ir_c^2\omega}
{\kappa_c(r_cu-1)^2}
-\frac{2ir_h^2\omega}
{\kappa_h(r_hu-1)^2}
-\frac{4ir_c\omega}
{\kappa_cu-\kappa_cr_cu^2}
\nonumber\\
&&
-\frac{4ir_h\omega}
{\kappa_hu-\kappa_hr_hu^2}
+\frac{2r_c^2\omega^2}
{\kappa_c^2(r_cu-1)^2}
+\frac{2r_h^2\omega^2}
{\kappa_h^2(r_hu-1)^2}
+\frac{4r_cr_h\omega^2}
{\kappa_c\kappa_h(r_cu-1)(r_hu-1)}
\nonumber\\
&&
-i\omega
\left(
\frac{r_c}
{\kappa_cA(u)-\kappa_cr_cuA(u)}
+\frac{r_h}
{\kappa_hA(u)-\kappa_hr_huA(u)}
\right)A'(u)
-\frac{A'(u)^2}{A(u)^2}
+4r_h^2\phi'(u)^2
\nonumber\\
&&
-\frac{ir_c\omega B'(u)}
{\kappa_cB(u)-\kappa_cr_cuB(u)}
-\frac{ir_h\omega B'(u)}
{\kappa_hB(u)-\kappa_hr_huB(u)}
+\frac{1}{A(u)}
\left(
\frac{3A'(u)}{u}+2A''(u)
\right)
\Bigg).
\end{eqnarray}
Based on $\lambda_\mathrm{0}$ and $s_\mathrm{0}$, the perturbed equation \eqref{eqschieq} can be solved numerically by using the improved $\mathrm{AIM}$ \cite{Cho:2009cj}. 
Following the same procedure described above, one can systematically derive the corresponding coefficient functions $\lambda_0$ and $s_0$ for the electromagnetic field perturbation. Given their highly cumbersome and lengthy analytical forms, we omit presenting them explicitly here for the sake of brevity. Nevertheless, these analytical expressions are exactly implemented in our numerical codes to compute the quasinormal frequencies for the electromagnetic sector.

\subsection{WKB method}

To analyze the spectral properties in the frequency domain, we utilize the semi-analytic WKB approximation technique \cite{Iyer:1986np,Konoplya:2011qq}. This approach is particularly effective for scattering problems involving potential barriers and has been extensively adapted for black hole physics.
This method, first proposed by Schutz and Will, was used to address black hole scattering problems \cite{Kokkotas:1988fm}. Later, further developments were made by Iyer, Will, and Konoplya  \cite{Konoplya:2011qq}. In this paper, we consider the most commonly used sixth-order WKB approximation method \cite{Konoplya:2011qq}
\begin{align}
\frac{i(\omega^2 - V_0)}{\sqrt{-2V_0''}} - \sum_{i=2}^6 \Lambda_i = n + \frac{1}{2}, \quad (n = 0, 1, 2, \cdots)
\end{align}
where \( V''(r_0) \) is the value of the second derivative of the effective potential with respect to \( r_* \) at its maximum point \( r_0 \) defined by the solution of the equation \( \left. \frac{dV}{dr_*} \right|_{r_*=r_0}= 0 \). \( V_0 \) represents the maximum value of the effective potential, and \( \Lambda_i \) is the \( i \)th-order revision terms depending on the values of the effective potential. This semi-analytical method has been applied extensively in numerous black hole spacetime cases. 
It should be pointed out here the WKB approach works well for situations where the multipole number is larger compared to the overtone: $l\geq n$, the WKB approach produces less accurate outcomes for $l<n$ \cite{Iyer:1986nq,Konoplya:2003ii}. 

By analyzing these errors, we provide a detailed assessment of the accuracy of our QNM estimates. 
We further consider the percentage deviation $\Delta_{AW}$ of QNMs obtained via the AIM and WKB methods. The relative error $\Delta_{AW}$ between two methods is defined by
\begin{eqnarray}
\Delta_{AW}=\frac{|\omega_{AIM}-\omega_{WKB}|}{|\omega_{AIM}|}\times 100\%.
\end{eqnarray}

\subsection{Numerical results}

Tables~\ref{bb1}, \ref{bb2}, and \ref{bb3} summarize the fundamental ($n=0$) quasinormal frequencies for both axial gravitational and electromagnetic perturbations. The results derived from the AIM and WKB approaches show excellent agreement. The relative error $\Delta_{AW}$ is typically well below $0.1\%$, confirming the reliability of our numerical data.

\begin{figure}[H]
\centering
\subfigure[$\epsilon=-1$]{\label{fig7-1} 
\includegraphics[width=3.1in]{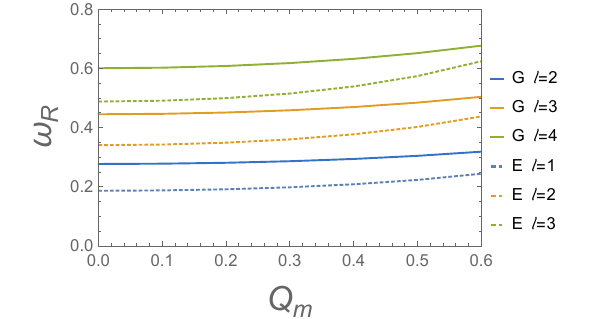}}
\subfigure[$\epsilon=-1$]{\label{fig7-2} 
\includegraphics[width=3.05in]{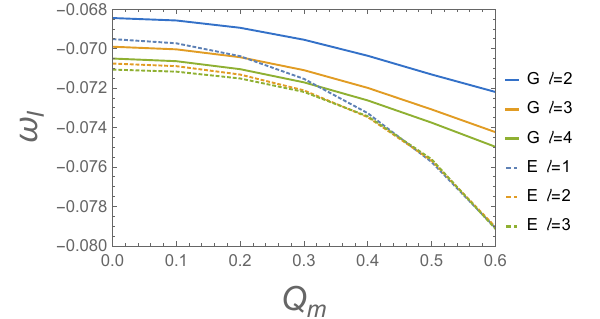}}
\subfigure[$\epsilon=0 $]{\label{fig7-3} 
\includegraphics[width=3.1in]{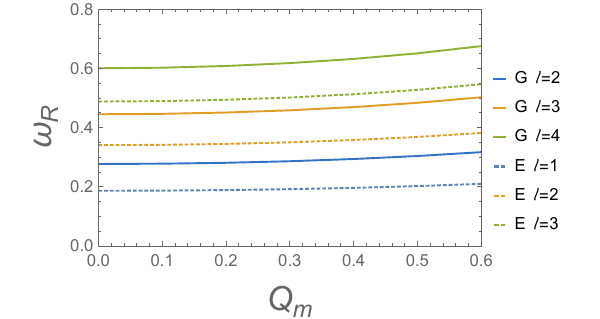}}
\subfigure[$\epsilon=0$]{\label{fig7-4} 
\includegraphics[width=3.05in]{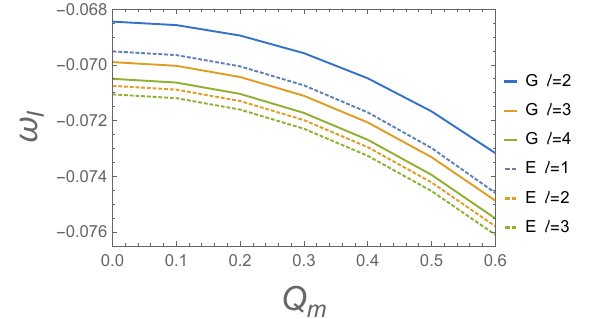}}
\subfigure[$\epsilon=1 $]{\label{fig7-5} 
\includegraphics[width=3.1in]{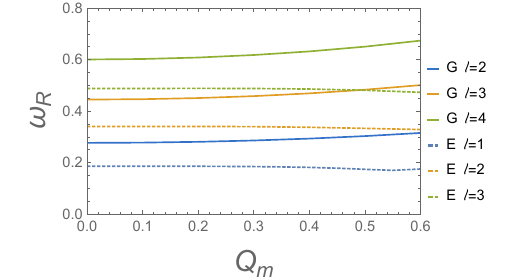}}
\subfigure[$\epsilon=1$]{\label{fig7-6} 
\includegraphics[width=3.05in]{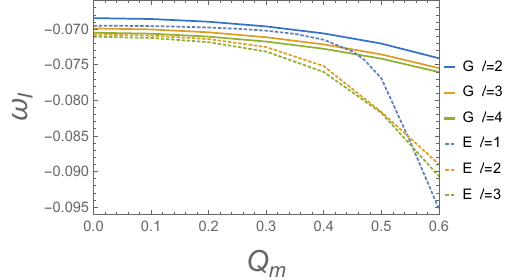}}
\caption{Variation of the gravitational and electromagnetic fundamental quasinormal frequencies ($n=0$) with respect to the magnetic charge $Q_{\mathrm{m}}$ for a black hole with fixed mass $M=1$ and cosmological constant $\Lambda=0.05$. Panels (a) and (b) correspond to $\epsilon=-1$, (c) and (d) to $\epsilon=0$, and (e) and (f) to $\epsilon=1$. In each panel, the solid lines represent gravitational perturbations for $l=2,3,4$ and dashed lines represent electromagnetic perturbations for $l=1,2,3$. The real and imaginary parts generally increase with increasing $Q_{\mathrm{m}}$, except for the anomalous electromagnetic branch with $\epsilon=1$.}\label{fig7}
\end{figure}

We first analyze the impact of the magnetic charge $Q_{\text{m}}$, which is depicted in Table~\ref{bb1} and Fig.~\ref{fig7}. Both the real part of the frequency $\operatorname{Re}(\omega)$ and the magnitude of the imaginary part $|\operatorname{Im}(\omega)|$ generally increase with an increase in $Q_{\text{m}}$, except for the anomalous electromagnetic branch with $\epsilon=1$. $\operatorname{Re}(\omega)$ represents the oscillation frequency of the perturbation, while $|\operatorname{Im}(\omega)|$ represents the damping rate. This implies that a larger magnetic charge results in faster oscillations and a more rapid decay of the gravitational perturbation.
For the electromagnetic perturbation with $\epsilon=1$, the damping rate $|\operatorname{Im}(\omega)|$ rises sharply as $Q_{\text{m}}$ increases, which appears as the steep drop of the blue dashed curve in Fig.~\ref{fig7}(f). This behavior has the same origin as the anomalous trajectory in Fig.~\ref{fig9}(c). As shown in Fig.~\ref{fig5}(c), for $\epsilon=1$ the nonlinear coupling anomalously suppresses the effective potential barrier as $Q_{\text{m}}$ increases, in contrast to the $\epsilon\le 0$ cases where the barrier grows higher. The lowered barrier can no longer confine the electromagnetic mode efficiently, so its energy leaks away rapidly and the damping rate increases sharply.

\begin{table*}[h!]
\caption{Fundamental QNM frequencies for gravitational ($l=2$) and electromagnetic ($l=1$) perturbations with $M=1$ and $\Lambda=0.05$.}\label{bb1}
\resizebox{\linewidth}{!}{
\begin{tabular}{|c|c|c|c|c|c|c|c|} \hline
&  & \multicolumn{3}{|c|}{gravitational field perturbation} & \multicolumn{3}{|c|}{electromagnetic field perturbation}   \\ \hline
$\epsilon$&  $Q_{\text{m}}$ &  AIM  &  WKB  &   $\Delta_{AW}$   &  AIM  &  WKB  &   $\Delta_{AW}$            \\ \hline
\multirow{3}{*}{1}& 0.1  & $  0.278039 -0.0685675 i$  & $ 0.278044 - 0.0685307   i$ & $0.0129776\% $ & $   0.186374-0.0695683i$  & $ 0.186425 - 0.069367   i$   & $0.104353\%$      \\ 
 & 0.3 &$0.286367 -0.0695433i$ & $0.286370 - 0.0695128  i$   & $0.0104035\%$ &  $0.191863 - 0.0707426i$ & $0.191835 - 0.0708107  i$   &$0.036163\% $ \\ 
  & 0.6 &$  0.315539-0.0740853i$   & $  0.315289 - 0.0740364  i$   &$0.0785703\%$ &  $0.175600-0.0952443i$  & $0.171869 - 0.0953679 i$    & $1.86904\%$      \\ \hline
  \multirow{3}{*}{0}& 0.1  & $0.278040-0.0685672i$  & $0.278047 - 0.0685298 i$      &$0.013252\%$ & $0.187019 -0.0696448i$  & $0.187072 - 0.0696775 i$    &  $0.0309554\%$       \\ 
   & 0.3 & $0.286298-0.0695743i$  & $0.286304 - 0.0695469 i$    &  $0.00951721\%$ & $0.191895-0.0707322i$  & $0.191837 - 0.0708023 i$    & $0.0443915\%$\\ 
   & 0.6 &$   0.317066-0.0731668i$  & $ 0.316902 - 0.0731263  i$   &$0.0520156\%$& $0.209995 -0.0745952i$ & $0.209854 - 0.0745773 i$   & $0.0640342\%$ \\\hline
   \multirow{3}{*}{-1}& 0.1  & $0.2780410 - 0.0685668 i$  & $ 0.278046 - 0.068530 i$     & $0.0129747\%$  & $0.187658 -0.0697235i$ & $0.187608 - 0.0696539   i$   & $0.0429693\%$    \\ 
   & 0.3 & $0.286357-0.0695410i$ & $0.286369 - 0.0695128 i$    &  $0.0104439\%$ & $0.198184 -0.0715274i$ & $0.198605 - 0.0718025  i$   & $0.238861\%$\\ 
   & 0.6 &$ 0.318658-0.0721916i$   & $ 0.318555 - 0.0722471 i$   &$0.0355968\%$& $0.244062 -0.0790832i$  & $0.245523 - 0.0792050  i$    & $0.571574\%$ \\\hline
 \end{tabular}}
\end{table*}

\begin{table*}[h!]
\caption{Fundamental QNM frequencies for gravitational ($l=2$) and electromagnetic ($l=1$) field perturbations with $M=1$ and $Q_{\text{m}}=0.3$.}\label{bb2}
\resizebox{\linewidth}{!}{
\begin{tabular}{|c|c|c|c|c|c|c|c|} \hline
&  & \multicolumn{3}{|c|}{gravitational field perturbation} & \multicolumn{3}{|c|}{electromagnetic field perturbation}   \\ \hline
$\epsilon$&  $\Lambda$ &  AIM  &  WKB  &   $\Delta_{AW}$   &  AIM  &  WKB  &   $\Delta_{AW}$            \\ \hline
\multirow{3}{*}{1} & 0.03  & $0.327657 -0.0785675i$ & $0.327632 - 0.0785274  i$   & $0.0140466\%$ & $0.210259 -0.0796292i$  & $0.211394 - 0.0771965   i$   & $1.19393\%$        \\ 
& 0.05  & $0.286239 -0.0696076i$ & $0.286239 - 0.0695808 i$   & $0.00908052\%$&  $0.184983 -0.0702106i$ & $  0.18676 - 0.0703966 i$   &$0.903163\%$   \\ 
 & 0.07  & $0.237710-0.0586194i$  & $0.237708 - 0.0586065 i$    & $0.00537367\%$&  $ 0.154675 -0.0589155i$  & $0.155393 - 0.0592603 i$   &$0.481237\%$    \\ \hline
  \multirow{3}{*}{0} & 0.03  & $0.327718 -0.0785223i$  & $0.327708 - 0.0784760  i$    &  $0.014058\%$   & $0.218602 -0.0804673i$  & $ 0.218544 - 0.0805638 i$      &  $0.0483773\%$    \\ 
  & 0.05  & $0.286298-0.0695743i$  & $0.286304 - 0.0695469 i$    & $0.00951721\%$&  $0.191895-0.0707322i$ & $  0.191837 - 0.0708023 i$    & $0.0443898\%$\\ 
   & 0.07  & $0.237767-0.0585981i$  & $0.237768 - 0.0585858 i$    & $0.00505481\%$ &  $0.160118 -0.0591910i$  & $0.160162 - 0.0592045 i$   &$0.0272163\%$ \\\hline
   \multirow{3}{*}{-1}& 0.03  & $0.327779 -0.0784770i$ & $0.327785 - 0.0784244  i$    &  $0.0156842\%$  & $0.226133 -0.0817082 i$  & $0.226549 - 0.0807616 i$     & $0.430093\%$    \\ 
   & 0.05  & $0.286357-0.0695410i$ & $0.286369 - 0.0695128  i$   & $0.0104439\%$  & $0.198184 -0.0715274i$ & $ 0.198557 - 0.0717788  i$    &$0.213585\%$\\ 
   & 0.07  & $0.237824 -0.0585768i$  & $0.237855 - 0.0585586 i$    & $0.0144856\%$  & $0.165104-0.0596258i$  & $0.165241 - 0.0596831 i$   &$0.0846287\%$\\\hline
 \end{tabular}}
\end{table*}

\begin{table*}[h!]
\caption{Fundamental QNM frequencies for gravitational and electromagnetic  field perturbations with $M=1$, $Q_{\text{m}}=0.3$ and $\Lambda=0.05$.}\label{bb3}
\resizebox{\linewidth}{!}{
\begin{tabular}{|c|c|c|c|c|c|c|c|c|} \hline
& \multicolumn{4}{|c|}{gravitational field perturbation} & \multicolumn{4}{|c|}{electromagnetic field perturbation}  \\ \hline
$\epsilon$ &  \multicolumn{1}{|c|}{$l$}&\multicolumn{1}{|c|}{AIM} & \multicolumn{1}{|c|}{WKB} & \multicolumn{1}{|c|}{$\Delta_{AW}$}  & \multicolumn{1}{|c|}{$l$}&\multicolumn{1}{|c|}{AIM} & \multicolumn{1}{|c|}{WKB} & \multicolumn{1}{|c|}{$\Delta_{AW}$} \\ \hline
\multirow{3}{*}{1} & 2  & $0.286239-0.0696076i$ & $0.286239 - 0.0695808 i$   & $0.00908052\%$    & $1$ & $0.1849830 -0.0702106i$ & $0.186760 - 0.0703966 i$ &$0.903163\%$ \\
 & 3  & $0.458477-0.0711285i$ & $0.458476 - 0.0711286 i$   & $0.0000314192\%$   & $2$ & $0.3400230 -0.0725128i$ & $0.339923 - 0.0724513 i$ &$0.033749\%$ \\
 & 4  & $0.618096-0.0717398i$  & $0.618096 - 0.0717398 i$    & $0.0000107106\%$ & $3$ & $0.488152-0.0731401i$ & $0.488087 - 0.0731173 i$ &  $0.0139642\%$\\ \hline
\multirow{3}{*}{0}  & 2  & $0.286298 -0.0695743i$  & $0.286304 - 0.0695469 i$    &  $0.00951721\%$ & $1$ & $0.191895 -0.0707322i$ & $0.191837 - 0.0708023 i$ & $0.0443898\%$\\
 & 3  & $0.458532 -0.0711072i$  & $0.458532 - 0.0711073 i$    & $0.0000382967\%$& $2$ & $0.350330 -0.0719815i$ & $0.350324 - 0.0719833 i$ &$0.00168055\%$ \\
 & 4  & $0.618157 -0.0717218i$ & $0.618157 - 0.0717218 i$   & $0.0000106829\%$ & $3$ & $0.501798 -0.0722913i$ & $0.501795 - 0.0722911 i$  & $0.000646973\%$\\ \hline
\multirow{3}{*}{-1} & 2  & $0.286357 -0.0695410i$ & $0.286369 - 0.0695128 i$    &  $0.0104439\%$ & $1$ & $0.198184 -0.0715274i$ & $0.19785 - 0.0720767 i$  &$0.305108\%$ \\
 & 3  & $0.458588 -0.0710858i$ & $0.458588 - 0.071086  i$   & $0.0000716413\%$ & $2$ & $0.360198 -0.0721124i$ & $0.360291 - 0.0721088 i$ &$0.0254766\%$ \\
& 4  & $0.618217 -0.0717038i$  & $0.618217 - 0.0717038 i$    & $0.0000123430\%$  & $3$ & $0.515174 -0.0721906i$ & $0.515222 - 0.0722027 i$ &$0.00946616\%$ \\ \hline
\end{tabular}}
\end{table*}

While the magnetic charge governs the non-trivial spectral shifts shown above, the response of the QNMs to the remaining background parameters follows standard, monotonic trends. We therefore report these results in tabular form. 
Table~\ref{bb2} lists the frequencies for varying cosmological constant $\Lambda$. Across all branches ($\epsilon = 0, \pm 1$), both $\text{Re}(\omega)$ and $|\text{Im}(\omega)|$ decrease steadily as $\Lambda$ grows. This slower oscillation and extended mode lifetime can be understood geometrically: an increasing $\Lambda$ shrinks the spatial cavity between the event horizon $r_h$ and the cosmological horizon $r_c$, effectively forming a tighter resonant waveguide that favors lower-frequency, longer-lived modes.

Similarly, Table~\ref{bb3} summarizes the influence of the multipole number $l$. As is typical for black hole perturbations, $\operatorname{Re}(\omega)$ grows substantially with $l$, whereas the damping rate $|\operatorname{Im}(\omega)|$ registers only a marginal increase. This reflects the familiar physics of the centrifugal barrier: higher angular momentum raises the effective potential height, driving up the oscillation frequency while leaving the decay timescale largely unaffected.

\begin{figure}[H]
\centering
\subfigure[$\epsilon=-1$]{\label{fig8-1} 
\includegraphics[width=2.1in]{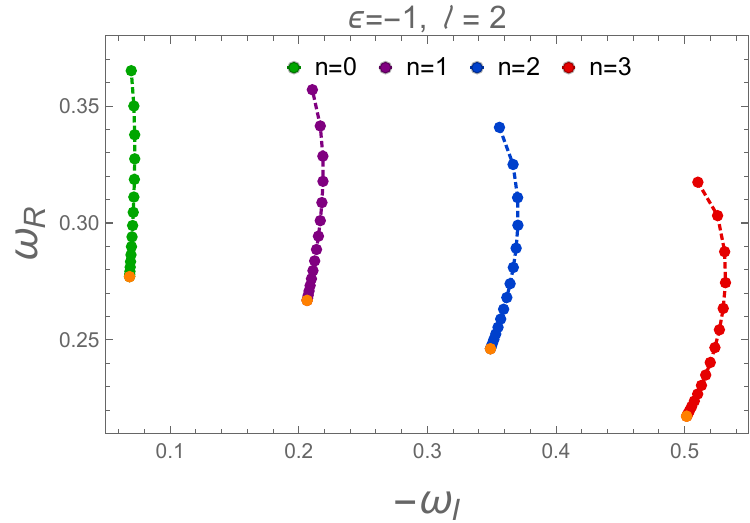}}
\hfill
\subfigure[$\epsilon=0$]{\label{fig8-2} 
\includegraphics[width=2.05in]{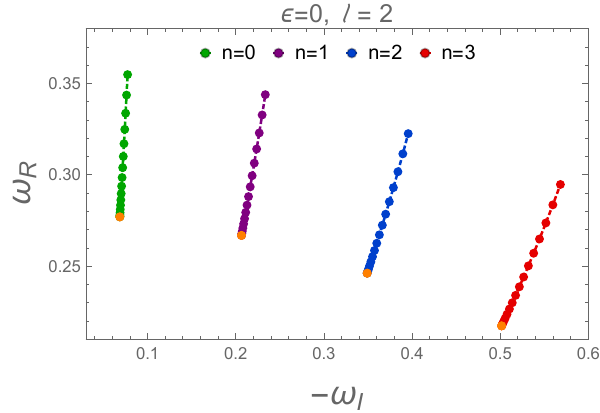}}
\hfill
\subfigure[$\epsilon=1$]{\label{fig8-3} 
\includegraphics[width=2.05in]{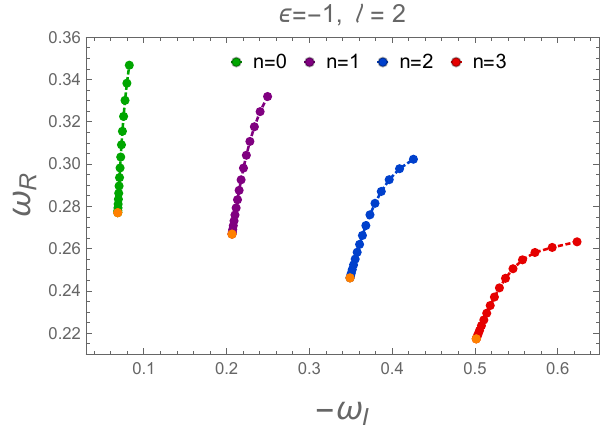}}
\caption{Behaviors of the real and imaginary parts of the quasinormal frequencies under gravitational perturbations for varying magnetic charge $Q_{\text{m}} \in [0, 0.6]$. The parameters are set to $l=2$, $M=1$, and $\Lambda=0.05$. The orange points denote $Q_{\text{m}}=0$, corresponding to the limit where the spacetime reduces to an uncharged Schwarzschild-de Sitter black hole. Different colored curves represent the QNMs for different overtone numbers $n$.}\label{fig8}
\end{figure}

\begin{figure}[H]
\centering
\subfigure[$\epsilon=-1$]{\label{fig9-1} 
\includegraphics[width=2.1in]{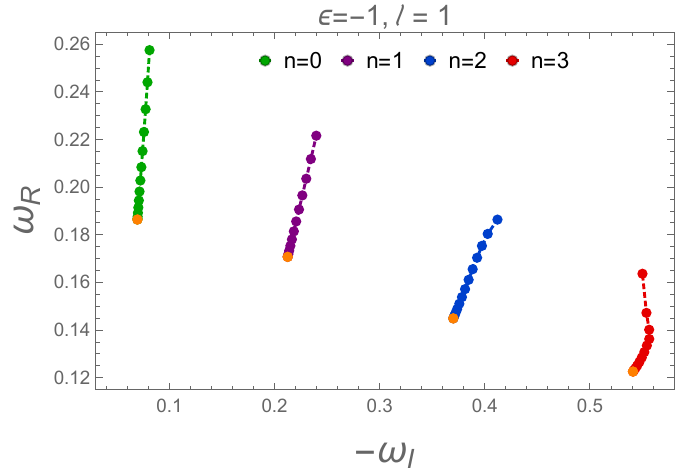}}
\hfill
\subfigure[$\epsilon=0$]{\label{fig9-2} 
\includegraphics[width=2.05in]{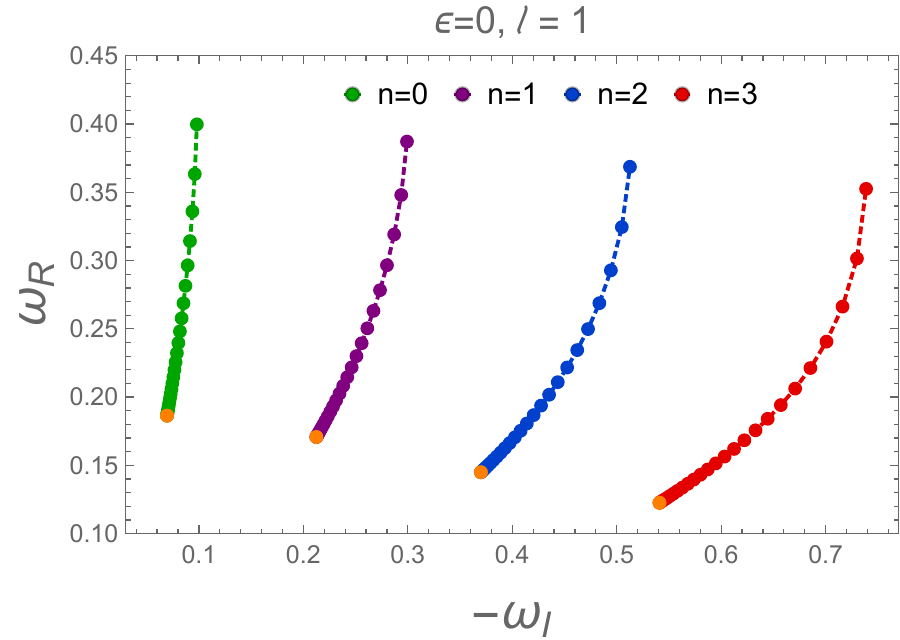}}
\hfill
\subfigure[$\epsilon=1$]{\label{fig9-3} 
\includegraphics[width=2.05in]{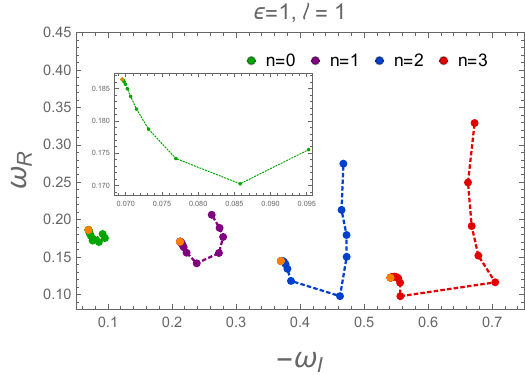}}
\caption{Behaviors of the real and imaginary parts of the quasinormal frequencies under electromagnetic perturbations for varying magnetic charge $Q_{\text{m}}$. The parameters are fixed at $l=1$, $M=1$, and $\Lambda=0.05$. The orange points denote $Q_{\text{m}}=0$, corresponding to the uncharged Schwarzschild-de Sitter limit. Different colored curves represent the QNMs for different overtone numbers $n$.}\label{fig9}
\end{figure}

To further illustrate the spectral properties in the complex frequency plane, we plot the trajectories of the real ($\omega_R$) and imaginary ($-\omega_I$) parts of the quasinormal frequencies in Figs. \ref{fig8} and \ref{fig9}. Fig. \ref{fig8}  depicts the gravitational sector ($l=2$), while Fig. \ref{fig9} represents the electromagnetic sector ($l=1$), both showing the evolution of the modes as the magnetic charge $Q_{\text{m}}$ varies continuously. For all branches of $\epsilon$ and different overtone numbers ($n=0, 1, 2, 3$), the variation of $Q_{\text{m}}$ induces a clear migration of the frequencies. Notably, as $Q_{\text{m}} \to 0$ (indicated by the orange points), all curves smoothly converge to the uncharged Schwarzschild-de Sitter (SdS) limit, confirming the consistency of our numerical framework. Furthermore, it is evident that higher overtones (larger $n$) generally correspond to significantly larger damping rates ($-\omega_I$), while their oscillation frequencies ($\omega_R$) remain relatively clustered.

A particularly interesting feature emerges in the electromagnetic sector for $\epsilon = 1$, as depicted in Fig. \ref{fig9-3}. Unlike the monotonic or smoothly curving trajectories observed in other parameter spaces, the QNM curves in Fig. \ref{fig9-3} exhibit a highly anomalous topological shape, bending irregularly as $Q_{\text{m}}$ increases. This distinct spectral behavior in the complex plane is profoundly intertwined with the underlying background geometry. By referring back to the effective potential for the electromagnetic perturbation in Fig. \ref{fig5-3}, one can observe that the behavior of the potential barrier's peak under variations of $Q_{\text{m}}$ deviates significantly from the patterns seen in Figs. \ref{fig5-1} and \ref{fig5-2}. The abnormal suppression and shift of the potential barrier directly alter the scattering cavity, thereby manifesting as the unconventional frequency trajectories observed in Fig. \ref{fig9-3}. This correspondence elegantly highlights how sensitive the ringing of the black hole is to the morphology of its effective potential under nonlinear electrodynamic corrections.

\begin{figure}[!htbp]
\centering
\subfigure[$\epsilon=-1$]{\label{fig10-1} 
\includegraphics[width=2.1in]{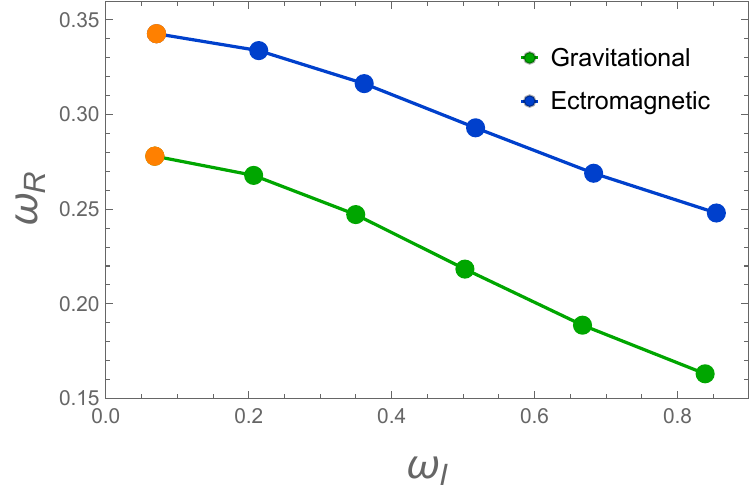}}
\hfill
\subfigure[$\epsilon=0$]{\label{fig10-2} 
\includegraphics[width=2.05in]{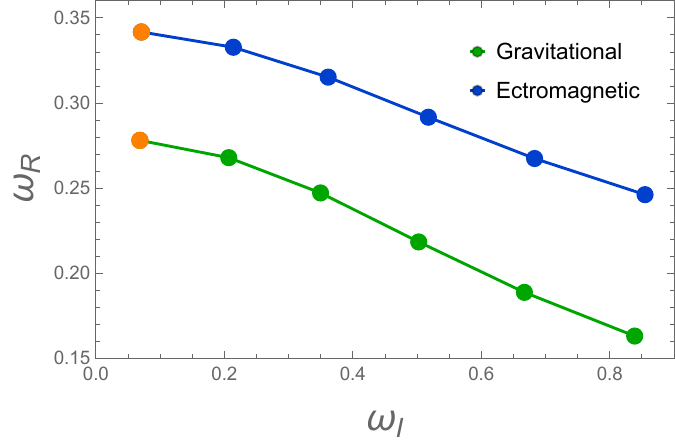}}
\hfill
\subfigure[$\epsilon=1$]{\label{fig10-3} 
\includegraphics[width=2.05in]{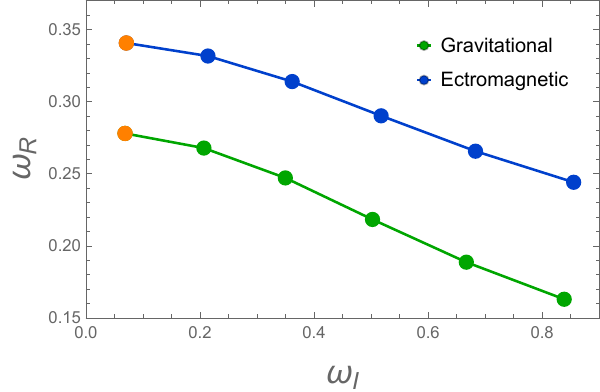}}
\caption{Quasinormal frequencies in the complex plane for different overtone numbers $n$ ($n=0,1,\dots,5$). The green curves represent gravitational perturbations with $l=2$, while the blue curves represent electromagnetic perturbations with $l=1$. The orange points specifically highlight the fundamental modes ($n=0$). Panels (a), (b), and (c) correspond to $\epsilon=-1$, $\epsilon=0$, and $\epsilon=1$, respectively. In all cases, the parameters are set to $M=1$, $Q_{\text{m}}=0.1$, and $\Lambda=0.05$.}\label{fig10}
\end{figure}

Finally, in Fig. \ref{fig10}, we fix the magnetic charge at $Q_{\text{m}} = 0.1$ to directly compare the overtone spectra ($n = 0, 1, \dots, 5$) between the gravitational (green curves) and electromagnetic (blue curves) perturbations. The orange points specifically highlight the fundamental modes ($n=0$). The results show that as the overtone number $n$ increases, the damping rate $-\omega_I$ increases almost linearly for both fields, indicating that higher-order modes dissipate much faster. Meanwhile, the real part $\omega_R$ gradually decreases. Under the given parameters, the gravitational perturbations exhibit a higher oscillation frequency but a lower damping rate compared to the electromagnetic perturbations. Furthermore, the overall topological structure of the overtone spectra remains robust across different nonlinear coupling parameters ($\epsilon = -1, 0, 1$), implying that the hierarchy of the overtones is primarily governed by the intrinsic properties of the black hole rather than the specific details of the nonlinear electrodynamic coupling.

\section{The Greybody factors}
\label{sec5}

In this section, we extend our analysis to the greybody factors of the magnetically charged dS black hole. The greybody factor essentially quantifies the inherent scattering characteristics of the spacetime's effective potential, reflecting the probability that an incoming wave from the cosmological horizon can successfully tunnel through the potential barrier and be absorbed by the black hole. In the static region of the de Sitter spacetime, the scattering problem is defined between the black-hole event horizon and the cosmological horizon. The appropriate boundary conditions for this scenario are defined as:
\begin{eqnarray}
\psi&=&T(\omega)e^{-i\omega r_*}, \quad r_* \rightarrow -\infty,\\
\psi&=&e^{-i\omega r_*} + R(\omega)e^{i\omega r_*}, \quad r_* \rightarrow +\infty,
\end{eqnarray}
where $T(\omega)$ and $R(\omega)$ represent the transmission and reflection coefficients, respectively. 
 
The squared amplitude of the wave function denotes the probability density. When incident waves encounter the effective potential barrier, they bifurcate into reflected and transmitted components. The transmission coefficient, $|T(\omega)|^2$, represents the fraction of the wave that effectively penetrates the barrier, which is standardly referred to as the greybody factor. Based on the conservation of probability, $R(\omega)$ and $T(\omega)$ satisfy the relation:
\begin{eqnarray}
|R(\omega)|^2+|T(\omega)|^2=1.
\end{eqnarray}
To compute these coefficients, we employ the sixth-order WKB approximation method, which is highly reliable for evaluating quantum tunneling through potential barriers. The reflection and transmission probabilities are given by:
\begin{eqnarray}
&&|R(\omega)|^2 = \frac{1} {1 + e^{-2\pi i K(\omega)}} ,\nonumber\\
&&|T(\omega)|^2 =\frac{1} {1 + e^{2\pi i K(\omega)}}= 1 -|R(\omega)|^2,\label{TR}
\end{eqnarray}
where the parameter $K(\omega)$ is determined via the WKB formula:
\begin{eqnarray}
K(\omega)= \frac{i\left( \omega^2 - V(r_0) \right)}{\sqrt{-2V''(r_0)}} + \sum_{i=2}^6 \Lambda_i.
\end{eqnarray}
For more detailed derivations and applications of this method, one can refer to reviews such as \cite{Konoplya:2019ppy,Konoplya:2011qq,Konoplya:2019hlu,Gogoi:2023fow,Liu:2023kxd,Singh:2024nvx} and the references therein.

\begin{figure}[H]
\centering
\subfigure[$\epsilon=-1$]{\label{fig11-1} 
\includegraphics[width=2in]{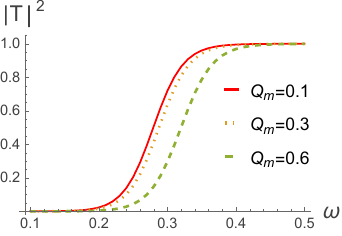}}
\hfill
\subfigure[$\epsilon=0$]{\label{fig11-2} 
\includegraphics[width=2in]{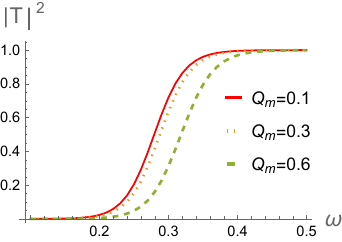}}
\hfill
\subfigure[$\epsilon=1$]{\label{fig11-3} 
\includegraphics[width=2in]{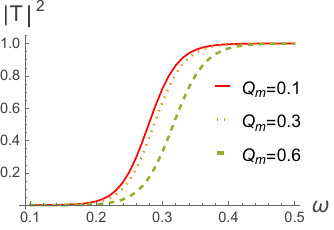}}
\hfill
\caption{The greybody factors for the gravitational field with $M=1$, $\Lambda=0.05$ and $l=2$.}\label{fig11}
\end{figure}

\begin{figure}[H]
\centering
\subfigure[\newtext{$\epsilon=-1$}]{\label{fig12-1} 
\includegraphics[width=2in]{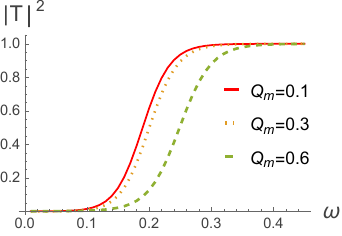}}
\hfill
\subfigure[$\epsilon=0$]{\label{fig12-2} 
\includegraphics[width=2in]{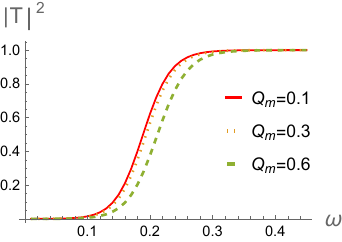}}
\hfill
\subfigure[\newtext{$\epsilon=1$}]{\label{fig12-3} 
\includegraphics[width=2in]{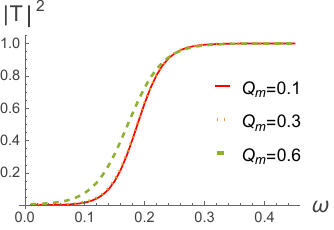}}
\hfill
\caption{The greybody factors for the electromagnetic  field with $M=1$, $\Lambda=0.05$ and $l=1$.}\label{fig12}
\end{figure}

In Figs.~\ref{fig11} and \ref{fig12}, we present the frequency-dependent greybody factors $|T(\omega)|^2$ for the gravitational ($l=2$) and electromagnetic ($l=1$) perturbations, respectively, under varying magnetic charge $Q_{\text{m}}$. For the gravitational perturbation and for the electromagnetic perturbation with $\epsilon\le 0$, increasing the magnetic charge $Q_{\text{m}}$ shifts the transmission curve to the right and decreases the greybody factor at a fixed frequency. In these cases, a larger magnetic charge enhances the height and width of the effective potential barrier, requiring higher incident wave frequencies to overcome it. The electromagnetic perturbation with $\epsilon=1$ is an exception and is discussed below.
For the electromagnetic perturbation with $\epsilon=1$ (Fig.~\ref{fig12}(c)), the transmission curves are not nested in the same regular order as those in the other panels. This is consistent with our QNM analysis and originates from the behavior of the effective potential barrier shown in Fig.~\ref{fig5}(c). For $\epsilon\le 0$, increasing $Q_{\text{m}}$ raises the barrier, so the wave tunnels less easily and the transmission curve shifts to the right in a regular manner. At $\epsilon=1$, in contrast, the barrier is anomalously suppressed as $Q_{\text{m}}$ increases, which makes the tunnelling easier instead of harder. The transmission curve for the largest $Q_{\text{m}}$ therefore breaks the ordered right-shifting pattern and lies outside the main distribution.

\begin{figure}[H]
\centering
\subfigure[$\epsilon=-1$]{\label{fig13-1} 
\includegraphics[width=2in]{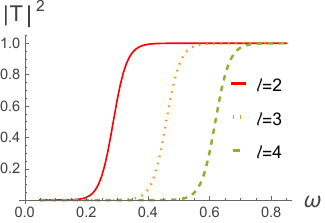}}
\hfill
\subfigure[$\epsilon=0$]{\label{fig13-2} 
\includegraphics[width=2in]{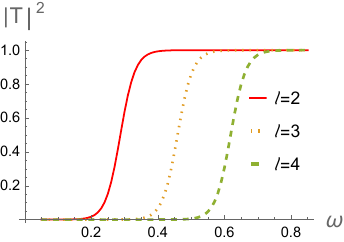}}
\hfill
\subfigure[$\epsilon=1$]{\label{fig13-3} 
\includegraphics[width=2in]{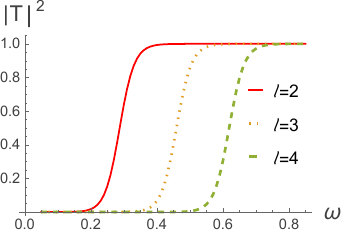}}
\hfill
\caption{The greybody factors for the gravitational field with $M=1$, $Q_{\text{m}}=0.3$ and  $\Lambda=0.05$.}\label{fig13}
\end{figure}

\begin{figure}[H]
\centering
\subfigure[$\epsilon=-1$]{\label{fig14-1} 
\includegraphics[width=2in]{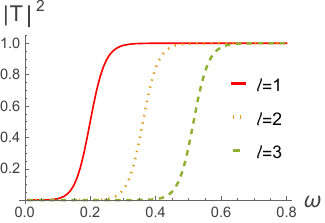}}
\hfill
\subfigure[$\epsilon=0$]{\label{fig14-2} 
\includegraphics[width=2in]{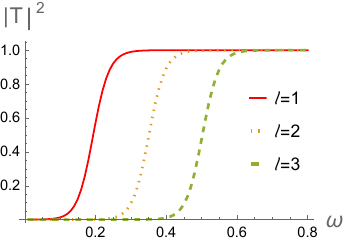}}
\hfill
\subfigure[$\epsilon=1$]{\label{fig14-3} 
\includegraphics[width=2in]{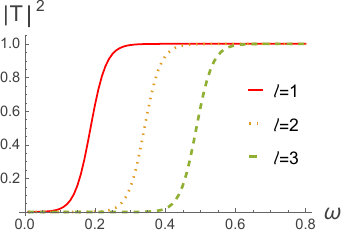}}
\hfill
\caption{The greybody factors for the electromagnetic  field with $M=1$, $Q_{\text{m}}=0.3$ and  $\Lambda=0.05$.}\label{fig14}
\end{figure}

Furthermore, we investigate the influence of the multipole number $l$ on the greybody factors, with the results for gravitational and electromagnetic fields displayed in Figs.~\ref{fig13} and \ref{fig14}, respectively. It is evident that the transmission probability diminishes drastically as $l$ increases. This phenomenon is deeply rooted in the structural mechanics of the effective potential: a higher multipole number translates to a significantly stronger centrifugal barrier in the spacetime geometry. This elevated centrifugal barrier effectively shields the black hole, markedly reducing the tunneling probability for modes with larger angular momentum. These scattering behaviors are in perfect agreement with the morphological evolution of the effective potentials illustrated in Figs.~\ref{fig4} and \ref{fig6}, providing a self-consistent picture of wave dynamics in this nonlinear electrodynamic background ~\cite{Du:2025kcx}.

The greybody factor is closely related to the Hawking sparsity of black hole radiation, as discussed in Refs.~\cite{Devi:2026tsr,Singh:2025fuv,Singh:2025tvk,Media:2025xpt,Singh:2024nvx}. A lower greybody factor indicates weaker transmission and therefore suggests sparser radiation. Thus, larger $Q_{\text{m}}$ and $l$ generally correspond to sparser radiation, while the electromagnetic sector with $\epsilon=1$ exhibits an anomalous response.

\section{Conclusion and discussion}
\label{sec6}

In this work, we studied axial gravitational and electromagnetic perturbations of magnetically charged black holes in a de Sitter background within string-inspired Euler-Heisenberg gravity. A key feature of the model is that the axial metric perturbations decouple completely from the electromagnetic sector, which enables the two perturbation equations to be reduced to separate Schr\"{o}dinger-like master equations. We further showed that the associated effective potentials remain positive throughout the parameter ranges considered. This supports the modal stability of the black hole under axial gravitational and electromagnetic perturbations.

Applying the AIM and 6th-order WKB methods, we computed the quasinormal frequencies and established excellent numerical agreement. Parametric analysis reveals that a larger cosmological constant $\Lambda$ yields slower, longer-lived modes, whereas higher multipole numbers $l$ predictably elevate the oscillation frequencies. Increasing the magnetic charge $Q_{\text{m}}$ generally drives faster oscillations and quicker decay. Notably, the electromagnetic sector under $\epsilon=1$ coupling exhibits anomalous, non-monotonic frequency trajectories in the complex plane—a direct signature of the deformed potential barrier induced by nonlinear electrodynamics. Overtone modes behave standardly, decaying more rapidly at higher $n$ with a spectral spacing robust against $\epsilon$ variations. Complementing the QNM spectra, the WKB evaluation of greybody factors demonstrates that wave transmission is generally suppressed for larger $Q_{\text{m}}$ or $l$, while the electromagnetic sector with $\epsilon=1$ exhibits an anomalous response associated with the deformation of the effective potential barrier.

Looking forward, exploring polar (even-parity) perturbations—where metric and gauge fluctuations are inherently coupled \cite{Nomura:2020tpc,Meng:2022oxg,Daghigh:2021psm}—remains an essential next step to complete the stability profile of magnetic black holes in string-inspired Euler-Heisenberg gravity.

 \vspace{1cm}
{\bf Acknowledgments}
 \vspace{0.5cm}

This research is supported by the National Natural Science Foundation of China (NNSFC) (Grant No.12365009), Natural Science Basic Research Program of Shaanxi Province (Program No. 2026JC-YBMS-0083) and Jiangxi Provincial Natural Science Foundation (No.20262BAC240347).

\bibliographystyle{unsrt}
\bibliography{reference}

@article{Heisenberg:1936nmg,
    author = "Heisenberg, W. and Euler, H.",
    title = "{Consequences of Dirac's theory of positrons}",
    journal = "Z. Phys.",
    volume = "98",
    number = "11-12",
    pages = "714--732",
    year = "1936",
    doi = "10.1007/BF01343663",
    eprint = "physics/0605038",
    archivePrefix = "arXiv",
    primaryClass = "physics"
}

@article{Obukhov:2002xa,
    author = "Obukhov, Y. N. and Rubilar, G. F.",
    title = "{Fresnel analysis of the wave propagation in nonlinear electrodynamics}",
    journal = "Phys. Rev. D",
    volume = "66",
    pages = "024042",
    year = "2002",
    doi = "10.1103/PhysRevD.66.024042",
    eprint = "gr-qc/0204028",
    archivePrefix = "arXiv",
    primaryClass = "gr-qc"
}

@article{Yajima:2000kw,
    author = "Yajima, H. and Tamaki, T.",
    title = "{Black hole solutions in Euler-Heisenberg theory}",
    journal = "Phys. Rev. D",
    volume = "63",
    pages = "064007",
    year = "2001",
    doi = "10.1103/PhysRevD.63.064007",
    eprint = "gr-qc/0005016",
    archivePrefix = "arXiv",
    primaryClass = "gr-qc"
}

@article{Ruffini:2013hia,
    author = "Ruffini, R. and Wu, Y. B. and Xue, S. S.",
    title = "{Einstein-Euler-Heisenberg Theory and charged black holes}",
    journal = "Phys. Rev. D",
    volume = "88",
    pages = "085004",
    year = "2013",
    doi = "10.1103/PhysRevD.88.085004",
    eprint = "1307.4951",
    archivePrefix = "arXiv",
    primaryClass = "hep-th"
}

@article{Breton:2019arv,
    author = "Bret{\'o}n, N. and L{\"a}mmerzahl, C. and Mac{\'\i}as, A.",
    title = "{Rotating black holes in the Einstein--Euler--Heisenberg theory}",
    journal = "Class. Quant. Grav.",
    volume = "36",
    number = "23",
    pages = "235022",
    year = "2019",
    doi = "10.1088/1361-6382/ab5169"
}

@article{Amaro:2022yew,
    author = "Amaro, D. and Breton, N. and L{\"a}mmerzahl, C. and Mac{\'\i}as, A.",
    title = "{Thermodynamics of the Einstein-Euler-Heisenberg rotating black hole}",
    journal = "Phys. Rev. D",
    volume = "105",
    number = "10",
    pages = "104046",
    year = "2022",
    doi = "10.1103/PhysRevD.105.104046"
}

@article{Guerrero:2020uhn,
    author = "Guerrero, M. and Rubiera-Garcia, D.",
    title = "{Nonsingular black holes in nonlinear gravity coupled to Euler-Heisenberg electrodynamics}",
    journal = "Phys. Rev. D",
    volume = "102",
    number = "2",
    pages = "024005",
    year = "2020",
    doi = "10.1103/PhysRevD.102.024005",
    eprint = "2005.08828",
    archivePrefix = "arXiv",
    primaryClass = "gr-qc"
}

@article{Nashed:2021ctg,
    author = "Nashed, G. G. L. and Nojiri, S.",
    title = "{Mimetic Euler-Heisenberg theory, charged solutions, and multihorizon black holes}",
    journal = "Phys. Rev. D",
    volume = "104",
    number = "4",
    pages = "044043",
    year = "2021",
    doi = "10.1103/PhysRevD.104.044043",
    eprint = "2107.13550",
    archivePrefix = "arXiv",
    primaryClass = "gr-qc"
}

@article{Bakopoulos:2024hah,
    author = "Bakopoulos, A. and Karakasis, T. and Mavromatos, N. E. and Nakas, T. and Papantonopoulos, E.",
    title = "{Exact black holes in string-inspired Euler-Heisenberg theory}",
    journal = "Phys. Rev. D",
    volume = "110",
    number = "2",
    pages = "024014",
    year = "2024",
    doi = "10.1103/PhysRevD.110.024014",
    eprint = "2402.12459",
    archivePrefix = "arXiv",
    primaryClass = "hep-th"
}

@article{Yasir:2025npe,
    author = "Yasir, M. and Mushtaq, F. and Tiecheng, X. and Javed, F.",
    title = "{Investigating the effects of particle motion and gravitational lensing of black hole in string-inspired Euler--Heisenberg theory}",
    journal = "Phys. Dark Univ.",
    volume = "48",
    pages = "101838",
    year = "2025",
    doi = "10.1016/j.dark.2025.101838"
}

@article{Vachher:2024ezs,
    author = "Vachher, Amnish and Islam, Shafqat Ul and Kumar Walia, Rahul and Ghosh, Sushant G.",
    title = "{Testing strong gravitational lensing effects of supermassive black holes with string-inspired metric: Observational signatures and EHT constraints}",
    eprint = "2405.06501",
    archivePrefix = "arXiv",
    primaryClass = "gr-qc",
    doi = "10.1016/j.aop.2025.170084",
    journal = "Annals Phys.",
    volume = "480",
    pages = "170084",
    year = "2025"
}

@article{Huang:2025jfa,
    author = "Huang, Hyat and Xu, Yukun and Lai, Meng-Yun and Zou, De-Cheng",
    title = "{Distinguishing black holes from string-inspired Euler{\textendash}Heisenberg theory through shadow images}",
    doi = "10.1142/S0218271825500634",
    journal = "Int. J. Mod. Phys. D",
    volume = "34",
    number = "13",
    pages = "2550063",
    year = "2025"
}

@article{Jiang:2024njc,
    author = "Jiang, Y. H. and Wang, T.",
    title = "{Accretion disks around magnetically charged black holes in string theory with an Euler-Heisenberg correction}",
    journal = "Phys. Rev. D",
    volume = "110",
    number = "10",
    pages = "103009",
    year = "2024",
    doi = "10.1103/PhysRevD.110.103009",
    eprint = "2408.10150",
    archivePrefix = "arXiv",
    primaryClass = "gr-qc"
}

@article{Berti:2007,
    author = "Berti, E. and Cardoso, V. and Gonzalez, J. A. and Sperhake, U.",
    title = "{Mining information from binary black hole mergers: A Comparison of estimation methods for complex exponentials in noise}",
    journal = "Phys. Rev. D",
    volume = "75",
    pages = "124017",
    year = "2007",
    doi = "10.1103/PhysRevD.75.124017"
}

@article{Nollert:1999,
    author = "Nollert, H. P. and Price, R. H.",
    title = "{Quantifying excitations of quasinormal mode systems}",
    journal = "J. Math. Phys.",
    volume = "40",
    pages = "980",
    year = "1999",
    doi = "10.1063/1.532698"
}

@article{Berti:2006,
    author = "Berti, E. and Cardoso, V. and Will, C. M.",
    title = "{On gravitational-wave spectroscopy of massive black holes with the space interferometer LISA}",
    journal = "Phys. Rev. D",
    volume = "73",
    pages = "064030",
    year = "2006",
    doi = "10.1103/PhysRevD.73.064030"
}

@article{Berti:2007b,
    author = "Berti, E. and Cardoso, J. and Cardoso, V. and Cavaglia, M.",
    title = "{Matched-filtering and parameter estimation of ringdown waveforms}",
    journal = "Phys. Rev. D",
    volume = "76",
    pages = "104044",
    year = "2007",
    doi = "10.1103/PhysRevD.76.104044"
}

@article{Isi:2019,
    author = "Isi, M. and Giesler, M. and Farr, W. M. and Scheel, M. A. and Teukolsky, S. A.",
    title = "{Testing the no-hair theorem with GW150914}",
    journal = "Phys. Rev. Lett.",
    volume = "123",
    pages = "111102",
    year = "2019",
    doi = "10.1103/PhysRevLett.123.111102"
}

@article{Zhao:2022gxl,
    author = "Zhao, Y. and Ren, X. and Ilyas, A. and Saridakis, E. N. and Cai, Y. F.",
    title = "{Quasinormal modes of black holes in f(T) gravity}",
    journal = "JCAP",
    volume = "10",
    pages = "087",
    year = "2022",
    doi = "10.1088/1475-7516/2022/10/087",
    eprint = "2204.11169",
    archivePrefix = "arXiv",
    primaryClass = "gr-qc"
}

@article{Jaramillo:2021,
    author = "Jaramillo, J. and Macedo, R. P. and Sheikh, L. A.",
    title = "{Pseudospectrum and Black Hole Quasinormal Mode Instability}",
    journal = "Phys. Rev. X",
    volume = "11",
    pages = "031003",
    year = "2021",
    doi = "10.1103/PhysRevX.11.031003"
}

@article{Cheung:2022,
    author = "Cheung, M. H. and Destounis, K. and Macedo, R. P. and Berti, E. and Cardoso, V.",
    title = "{Destabilizing the Fundamental Mode of Black Holes: The Elephant and the Flea}",
    journal = "Phys. Rev. Lett.",
    volume = "128",
    pages = "111103",
    year = "2022",
    doi = "10.1103/PhysRevLett.128.111103"
}

@article{Ishibashi:2003,
    author = "Ishibashi, A. and Kodama, H.",
    title = "{Stability of higher dimensional Schwarzschild black holes}",
    journal = "Prog. Theor. Phys.",
    volume = "110",
    pages = "901",
    year = "2003",
    doi = "10.1143/PTP.110.901"
}

@article{Wu:2018xza,
    author = "Wu, C.",
    title = "{Quasinormal frequencies of gravitational perturbation in regular black hole spacetimes}",
    journal = "Eur. Phys. J. C",
    volume = "78",
    number = "4",
    pages = "283",
    year = "2018",
    doi = "10.1140/epjc/s10052-018-5764-6"
}

@article{Yan:2020nvk,
    author = "Yan, Z. and Wu, C. and Guo, W.",
    title = "{Quasinormal modes of scalar field coupled to Einstein's tensor in the non-commutative geometry inspired black hole}",
    journal = "Nucl. Phys. B",
    volume = "973",
    pages = "115595",
    year = "2021",
    doi = "10.1016/j.nuclphysb.2021.115595",
    eprint = "2012.03004",
    archivePrefix = "arXiv",
    primaryClass = "nucl-th"
}

@article{Cho:2009cj,
    author = "Cho, H. T. and Cornell, A. S. and Doukas, J. and Naylor, W.",
    title = "{Black hole quasinormal modes using the asymptotic iteration method}",
    journal = "Class. Quant. Grav.",
    volume = "27",
    pages = "155004",
    year = "2010",
    doi = "10.1088/0264-9381/27/15/155004",
    eprint = "0912.2740",
    archivePrefix = "arXiv",
    primaryClass = "gr-qc"
}

@article{Cho:2011sf,
    author = "Cho, H. T. and Cornell, A. S. and Doukas, J. and Huang, T. R. and Naylor, W.",
    title = "{A New Approach to Black Hole Quasinormal Modes: A Review of the Asymptotic Iteration Method}",
    journal = "Adv. Math. Phys.",
    volume = "2012",
    pages = "281705",
    year = "2012",
    doi = "10.1155/2012/281705",
    eprint = "1111.5024",
    archivePrefix = "arXiv",
    primaryClass = "gr-qc"
}

@article{Iyer:1986np,
    author = "Iyer, S. and Will, C. M.",
    title = "{Black Hole Normal Modes: A WKB Approach. 1. Foundations and Application of a Higher Order WKB Analysis of Potential Barrier Scattering}",
    journal = "Phys. Rev. D",
    volume = "35",
    pages = "3621",
    year = "1987",
    doi = "10.1103/PhysRevD.35.3621"
}

@article{Iyer:1986nq,
    author = "Iyer, S.",
    title = "{Black Hole Normal Modes: A WKB Approach. 2. Schwarzschild Black Holes}",
    journal = "Phys. Rev. D",
    volume = "35",
    pages = "3632",
    year = "1987",
    doi = "10.1103/PhysRevD.35.3632"
}

@article{Konoplya:2003ii,
    author = "Konoplya, R. A.",
    title = "{Quasinormal behavior of the d-dimensional Schwarzschild black hole and higher order WKB approach}",
    journal = "Phys. Rev. D",
    volume = "68",
    pages = "024018",
    year = "2003",
    doi = "10.1103/PhysRevD.68.024018",
    eprint = "gr-qc/0303052",
    archivePrefix = "arXiv",
    primaryClass = "gr-qc"
}

@article{Konoplya:2019ppy,
    author = "Konoplya, R. A. and Zinhailo, A. F.",
    title = "{Hawking radiation of non-Schwarzschild black holes in higher derivative gravity: a crucial role of grey-body factors}",
    journal = "Phys. Rev. D",
    volume = "99",
    number = "10",
    pages = "104060",
    year = "2019",
    doi = "10.1103/PhysRevD.99.104060",
    eprint = "1904.05341",
    archivePrefix = "arXiv",
    primaryClass = "gr-qc"
}

@article{Konoplya:2011qq,
    author = "Konoplya, R. A. and Zhidenko, A.",
    title = "{Quasinormal modes of black holes: From astrophysics to string theory}",
    journal = "Rev. Mod. Phys.",
    volume = "83",
    pages = "793--836",
    year = "2011",
    doi = "10.1103/RevModPhys.83.793",
    eprint = "1102.4014",
    archivePrefix = "arXiv",
    primaryClass = "gr-qc"
}

@article{Kokkotas:1988fm,
    author = "Kokkotas, K. D. and Schutz, B. F.",
    title = "{Black Hole Normal Modes: A WKB Approach. 3. The Reissner-Nordstrom Black Hole}",
    journal = "Phys. Rev. D",
    volume = "37",
    pages = "3378--3387",
    year = "1988",
    doi = "10.1103/PhysRevD.37.3378"
}

@article{Konoplya:2019hlu,
    author = "Konoplya, R. A. and Zhidenko, A. and Zinhailo, A. F.",
    title = "{Higher order WKB formula for quasinormal modes and grey-body factors: recipes for quick and accurate calculations}",
    journal = "Class. Quant. Grav.",
    volume = "36",
    pages = "155002",
    year = "2019",
    doi = "10.1088/1361-6382/ab2e25",
    eprint = "1904.10333",
    archivePrefix = "arXiv",
    primaryClass = "gr-qc"
}

@article{Gogoi:2023fow,
    author = "Gogoi, D. J. and {\"O}vg{\"u}n, A. and Demir, D.",
    title = "{Quasinormal modes and greybody factors of symmergent black hole}",
    journal = "Phys. Dark Univ.",
    volume = "42",
    pages = "101314",
    year = "2023",
    doi = "10.1016/j.dark.2023.101314",
    eprint = "2306.09231",
    archivePrefix = "arXiv",
    primaryClass = "gr-qc"
}

@article{Oshita:2023cjz,
    author = "Oshita, N.",
    title = "{Greybody factors imprinted on black hole ringdowns: An alternative to superposed quasinormal modes}",
    journal = "Phys. Rev. D",
    volume = "109",
    number = "10",
    pages = "104028",
    year = "2024",
    doi = "10.1103/PhysRevD.109.104028",
    eprint = "2309.05725",
    archivePrefix = "arXiv",
    primaryClass = "gr-qc"
}

@article{Regge:1957td,
    author = "Regge, T. and Wheeler, J. A.",
    title = "{Stability of a Schwarzschild singularity}",
    journal = "Phys. Rev.",
    volume = "108",
    pages = "1063--1069",
    year = "1957",
    doi = "10.1103/PhysRev.108.1063"
}

@article{Zerilli:1970se,
    author = "Zerilli, F. J.",
    title = "{Effective potential for even parity Regge-Wheeler gravitational perturbation equations}",
    journal = "Phys. Rev. Lett.",
    volume = "24",
    pages = "737--738",
    year = "1970",
    doi = "10.1103/PhysRevLett.24.737"
}

@article{Nomura:2020tpc,
    author = "Nomura, K. and Yoshida, D. and Soda, J.",
    title = "{Stability of magnetic black holes in general nonlinear electrodynamics}",
    journal = "Phys. Rev. D",
    volume = "101",
    number = "12",
    pages = "124026",
    year = "2020",
    doi = "10.1103/PhysRevD.101.124026",
    eprint = "2004.07560",
    archivePrefix = "arXiv",
    primaryClass = "gr-qc"
}

@article{Meng:2022oxg,
    author = "Meng, K. and Zhang, S. J.",
    title = "{Gravito-electromagnetic perturbations and QNMs of regular black holes}",
    journal = "Class. Quant. Grav.",
    volume = "40",
    number = "19",
    pages = "195024",
    year = "2023",
    doi = "10.1088/1361-6382/acf3c6",
    eprint = "2210.00295",
    archivePrefix = "arXiv",
    primaryClass = "gr-qc"
}

@article{Daghigh:2021psm,
    author = "Daghigh, R. G. and Green, M. D.",
    title = "{Gravitational and electromagnetic radiation from an electrically charged black hole in general nonlinear electrodynamics}",
    journal = "Phys. Rev. D",
    volume = "105",
    number = "2",
    pages = "024055",
    year = "2022",
    doi = "10.1103/PhysRevD.105.024055",
    eprint = "2106.01412",
    archivePrefix = "arXiv",
    primaryClass = "gr-qc"
}

@inproceedings{Bonanno:2025ffy,
    author = "Bonanno, Alfio and Silveravalle, Samuele",
    title = "{Spontaneous ghostification: how a dying black hole comes back as a naked singularity}",
    booktitle = "{59th Rencontres de Moriond on Gravitation}: {Moriond 2025 Gravitation}",
    eprint = "2505.20360",
    archivePrefix = "arXiv",
    primaryClass = "gr-qc",
    month = "5",
    year = "2025"
}

@article{Zhang:2025xqt,
    author = "Zhang, Xufen and Zou, De-Cheng and Zhang, Chao-Ming and Zhang, Ming and Yue, Rui-Hong",
    title = "{Perturbations of massless external fields on magnetically charged black holes in string-inspired Euler-Heisenberg theory}",
    eprint = "2508.17736",
    archivePrefix = "arXiv",
    primaryClass = "gr-qc",
    doi = "10.1088/1674-1137/ade661",
    journal = "Chin. Phys. C",
    volume = "49",
    number = "10",
    pages = "105109",
    year = "2025"
}

@article{Zhang:2026nog,
    author = "Zhang, Ming and Chen, Guo-Xin and Zhang, Lei and Li, Sheng-Yuan and Zhang, Xufen and Zou, De-Cheng",
    title = "{Quasinormal modes and greybody factors of magnetically charged de Sitter black holes probed by massless external fields in Einstein{\textendash}Euler{\textendash}Heisenberg gravity}",
    eprint = "2603.09304",
    archivePrefix = "arXiv",
    primaryClass = "gr-qc",
    doi = "10.1088/1572-9494/ae42b1",
    journal = "Commun. Theor. Phys.",
    volume = "78",
    number = "5",
    pages = "055406",
    year = "2026"
}

@article{Zou:2025rbu,
    author = "Zou, De-Cheng and Zhang, Xufen and Zhang, Chao-Ming and Zhang, Ming and Yue, Rui-Hong",
    title = "{Axial gravitational quasinormal modes of magnetically charged black holes}",
    doi = "10.1088/1674-1137/adff01",
    journal = "Chin. Phys. C",
    volume = "49",
    number = "12",
    pages = "125109",
    year = "2025"
}

@article{Li:2026gqi,
    author = "Li, Sheng-Yuan and Myung, Yun Soo and Zhang, Ming and Zhang, Xufen and Zou, De-Cheng",
    title = "{Polar perturbations of dilaton-Euler{\textendash}Heisenberg black holes}",
    eprint = "2601.13521",
    archivePrefix = "arXiv",
    primaryClass = "gr-qc",
    doi = "10.1140/epjc/s10052-026-15836-4",
    journal = "Eur. Phys. J. C",
    volume = "86",
    number = "6",
    pages = "618",
    year = "2026"
}

@article{Blazquez-Salcedo:2016enn,
    author = "Bl{\'a}zquez-Salcedo, Jose Luis and Macedo, Caio F. B. and Cardoso, Vitor and Ferrari, Valeria and Gualtieri, Leonardo and Khoo, Fech Scen and Kunz, Jutta and Pani, Paolo",
    title = "{Perturbed black holes in Einstein-dilaton-Gauss-Bonnet gravity: Stability, ringdown, and gravitational-wave emission}",
    eprint = "1609.01286",
    archivePrefix = "arXiv",
    primaryClass = "gr-qc",
    doi = "10.1103/PhysRevD.94.104024",
    journal = "Phys. Rev. D",
    volume = "94",
    number = "10",
    pages = "104024",
    year = "2016"
}

@article{Aragon:2020xtm,
    author = "Arag{\'o}n, Almendra and Gonz{\'a}lez, P. A. and Papantonopoulos, Eleftherios and V{\'a}squez, Yerko",
    title = "{Quasinormal modes and their anomalous behavior for black holes in $f(R)$ gravity}",
    eprint = "2005.11179",
    archivePrefix = "arXiv",
    primaryClass = "gr-qc",
    doi = "10.1140/epjc/s10052-021-09193-7",
    journal = "Eur. Phys. J. C",
    volume = "81",
    number = "5",
    pages = "407",
    year = "2021"
}

@article{Cano:2021myl,
    author = "Cano, Pablo A. and Fransen, Kwinten and Hertog, Thomas and Maenaut, Simon",
    title = "{Gravitational ringing of rotating black holes in higher-derivative gravity}",
    eprint = "2110.11378",
    archivePrefix = "arXiv",
    primaryClass = "gr-qc",
    doi = "10.1103/PhysRevD.105.024064",
    journal = "Phys. Rev. D",
    volume = "105",
    number = "2",
    pages = "024064",
    year = "2022"
}

@article{Du:2025kcx,
    author = "Du, Yongbin and Sun, Jia-Rui and Zhang, Xiangdong",
    title = "{Information paradox and island of covariant black holes in LQG}",
    eprint = "2510.11921",
    archivePrefix = "arXiv",
    primaryClass = "gr-qc",
    doi = "10.1103/b1my-3v4r",
    journal = "Phys. Rev. D",
    volume = "113",
    number = "4",
    pages = "046017",
    year = "2026"
}

@article{Liu:2023kxd,
    author = "Liu, Yunlong and Zhang, Xiangdong",
    title = "{Quasinormal modes of Bardeen black holes with a cloud of strings*}",
    eprint = "2305.02642",
    archivePrefix = "arXiv",
    primaryClass = "gr-qc",
    doi = "10.1088/1674-1137/acf3d5",
    journal = "Chin. Phys. C",
    volume = "47",
    number = "12",
    pages = "125103",
    year = "2023"
}

@article{Lin:2024ubg,
    author = "Lin, Jianhui and Bravo-Gaete, Mois{\'e}s and Zhang, Xiangdong",
    title = "{Quasinormal modes, greybody factors, and thermodynamics of four dimensional AdS black holes in critical gravity}",
    eprint = "2401.02045",
    archivePrefix = "arXiv",
    primaryClass = "gr-qc",
    doi = "10.1103/PhysRevD.109.104039",
    journal = "Phys. Rev. D",
    volume = "109",
    number = "10",
    pages = "104039",
    year = "2024"
}

@article{Wu:2021pgf,
    author = "Wu, Xianglong and Zhang, Xiangdong",
    title = "{Connections between the Shadow Radius and the Quasinormal Modes of Kerr-Sen Black Hole}",
    eprint = "2112.11066",
    archivePrefix = "arXiv",
    primaryClass = "gr-qc",
    doi = "10.3390/universe8110604",
    journal = "Universe",
    volume = "8",
    number = "11",
    pages = "604",
    year = "2022"
}

@article{Devi:2026tsr,
    author = "Devi, Irengbam Roshila and Media, Ningthoujam and Singh, Yenshembam Priyobarta and Singh, Telem Ibungochouba",
    title = "{Strong Lensing and Quasinormal modes of black hole around global monopole}",
    eprint = "2604.05686",
    archivePrefix = "arXiv",
    primaryClass = "gr-qc",
    note = "arXiv:2604.05686 "
}

@article{Singh:2025fuv,
    author = "Singh, Yenshembam Priyobarta and Choudhury, Jayasri and Singh, Telem Ibungochouba and Gogoi, Dhruba Jyoti",
    title = "{Field perturbations and observables in a charged black hole within bumblebee gravity}",
    doi = "10.1140/epjp/s13360-025-07060-y",
    journal = "Eur. Phys. J. Plus",
    volume = "140",
    number = "11",
    pages = "1118",
    year = "2025"
}

@article{Singh:2025tvk,
    author = "Singh, Yenshembam Priyobarta and Media, Ningthoujam and Singh, Telem Ibungochouba",
    title = "{Strong lensing and Hawking spectra of charged black hole under Lorentz violation theory}",
    eprint = "2508.09864",
    archivePrefix = "arXiv",
    primaryClass = "gr-qc",
    doi = "10.1140/epjc/s10052-025-14929-w",
    journal = "Eur. Phys. J. C",
    volume = "85",
    number = "10",
    pages = "1223",
    year = "2025"
}

@article{Singh:2024nvx,
    author = "Singh, Yenshembam Priyobarta and Singh, Telem Ibungochouba",
    title = "{Greybody factor and quasinormal modes of scalar and Dirac field perturbation in Schwarzschild-de Sitter-like black hole in Bumblebee gravity model}",
    eprint = "2408.14945",
    archivePrefix = "arXiv",
    primaryClass = "hep-ph",
    doi = "10.1140/epjc/s10052-024-13627-3",
    journal = "Eur. Phys. J. C",
    volume = "84",
    number = "12",
    pages = "1245",
    year = "2024"
}

@article{Media:2025xpt,
    author = "Media, Ningthoujam and Singh, Y. Priyobarta and Laxmi, Y. Onika and Singh, T. Ibungochouba",
    title = "{Entropy correction and quasinormal modes of slowly rotating Kerr--Newman--de Sitter-like black hole in bumblebee gravity}",
    journal = "Gen. Relativ. Gravit.",
    volume = "57",
    pages = "80",
    year = "2025",
    doi = "10.1007/s10714-025-03414-0"
}

\end{document}